\documentclass[epj]{svjour}

\usepackage{hyperref}
\usepackage{epsfig,graphics,colordvi}
\usepackage{amsmath}
\usepackage{amssymb}
\usepackage{mciteplus}
\usepackage{dsfont}
\usepackage{cite}

\newcommand{\tr}{\operatorname{tr}}
\newcommand{\te}{\text}
\newcommand{\nn}{\nonumber}

\newcommand{\one}{\mathds{1}}
\newcommand{\eps}{\varepsilon}
\newcommand{\etapr}{\eta\hspace{0.05em}'}

\begin{document}

\title{Electromagnetic transitions in an effective chiral Lagrangian with the $\etapr$ and light
vector mesons}
\titlerunning{Electromagnetic transitions in an effective chiral Lagrangian}
\author{C.\ Terschl\"usen\inst{1} \and S.\ Leupold\inst{1} \and M.F.M.\ Lutz\inst{2}
}                     
%
%
\institute{Institutionen f\"or fysik och astronomi, Uppsala Universitet, Box 516, 75120 Uppsala, Sweden
\and GSI Helmholtzzentrum f\"ur Schwerionenforschung GmbH, Planckstra\ss e 1, 64291 Darmstadt, Germany}
\authorrunning{C.\ Terschl\"usen et al.}
\date{Received: \today / Revised version: date}
%
\abstract{
We consider the chiral Lagrangian with a nonet of Goldstone bosons and a nonet of light
vector mesons. The mixing between the pseudoscalar mesons $\eta$ and $\etapr$ is taken into account.
A novel counting scheme is suggested that is based on hadrogenesis, which conjectures a mass gap
in the meson spectrum of QCD in the limit of a large number of colors. Such a mass gap would justify to consider
the vector mesons and the $\etapr$ meson as light degrees of freedom.  The complete leading order
Lagrangian is constructed and discussed. As a first application it is tested against electromagnetic transitions
of light vector mesons to pseudoscalar mesons. Our parameters are determined by the experimental data on
photon decays of
the $\omega, \phi$ and $\etapr$ meson. In terms of such parameters we predict the
corresponding decays into virtual photons with either dielectrons or dimuons in the final state.
\PACS{
      {13.40.Gp}{Electromagnetic form factors}   \and
      {12.39.Fe}{Chiral Lagrangians} \and {12.40.Vv}{Vector-meson dominance}
     } 
} 
\maketitle

\section{Introduction}

An open challenge in hadron physics is the development of effective field theories describing strong interactions
in hadronic reactions and decays, see e.g.\ \cite{Weinberg:1978kz,Gasser:1983yg,Gasser:1984gg,Bando:1987br,Meissner:1987ge,Ecker:1988te,Ecker:1989yg,Jenkins:1995vb,Birse:1996hd,Kaiser:2000gs,Harada:2003jx,Bruns:2004tj,Lutz:2008km,Kampf:2009jh,Djukanovic:2010tb}.
For the energy region where only the Goldstone bosons participate, the dynamics can be successfully described
by chiral perturbation theory, however, only close to threshold
\cite{Weinberg:1978kz,Gasser:1983yg,Gasser:1984gg,Scherer:2002tk}.
For higher energies coupled-channel unitarity plays an increasingly important role and approaches that rest on
the chiral Lagrangian, but perform partial summation as to recover coupled-channel unitarity, are quite successful
(see e.g.
\cite{Lutz:2001yb,Lutz:2003fm,Danilkin:2010xd,Gasparyan:2010xz,Danilkin:2011fz}).
Though there is some dispute in the literature as to the rigor of such
approaches they appear to be highly predictive and grasp the right
physics.  Yet, for higher energies hadronic resonances like the
$\rho$, $\omega$, $K^\ast$ and $\phi$ mesons are known to be relevant
degrees of freedom.

Though it is well known how to incorporate more massive degrees of freedom into the chiral Lagrangian, it is unclear how
to organize systematic applications. The key issue is the identification of an optimal set of degrees of freedom in combination
with the construction of power counting rules. In a previous work two of the authors suggested a novel counting scheme for the
chiral Lagrangian which included the nonet of light vector mesons \cite{Lutz:2008km}. It is based on the hadrogenesis
conjecture and large-$N_c$ considerations \cite{Lutz:2001dr,Lutz:2003fm,Lutz:2004dz,Lutz:2005ip,Lutz:2007sk}.
Here and in the following $N_c$ denotes the number of colors \cite{'tHooft:1973jz,Witten:1979kh}.
So far the scheme suggested in \cite{Lutz:2008km} has been applied successfully
at tree level in  \cite{Leupold:2008bp,Terschluesen:2010ik}  and  in a coupled-channel
framework in \cite{Danilkin:2011fz}.
Whether it leads to a fully systematic effective field theory is an open issue.

Our current work should be seen in the context of the recently developed summation scheme \cite{Gasparyan:2010xz}.  
While at low energies a strict perturbative application of the chiral Lagrangian is applicable, as
the energy increases this is no longer true. In order to reach the resonance region an analytic extrapolation has been 
performed in \cite{Gasparyan:2010xz} that is stabilized by the
unitarity constraint. This leads to a controlled approximation of scattering amplitudes in the resonance region 
since the analytic continuation is based on conformal expansions that can be proven to converge uniformly in the 
resonance region. The important point is that such an 
extrapolation is possible with the knowledge of the subthreshold scattering amplitude, which is accessible in a conventional 
perturbative application of the effective Lagrangian.   

In this work we attempt a generalization and further systematization of the counting rules proposed in \cite{Lutz:2008km}.
The prime objective is the incorporation of the $\etapr$ meson and the derivation of a more transparent method to predict
for a given interaction the proper order in the power counting.
In particular we offer an explicit realization of the naturalness assumption for the counter terms
as is implied by the characteristic mass gap expected in the hadrogenesis picture
\cite{Lutz:2001dr,Lutz:2003fm,Lutz:2004dz,Lutz:2005ip,Lutz:2007sk}.
The generalized counting rules will be tested against hadronic and electromagnetic two- and three-body decays
of the pseudoscalar and vector mesons. After a successful determination of the relevant parameter
set we provide quantitative predictions for the processes $\omega \rightarrow \pi^0 l^+ l^-$, $\omega \rightarrow \eta\, l^+ l^-$
and $\phi \rightarrow \eta \,l^+ l^-$ with $l = \mu, e$ and the processes $\phi \rightarrow \etapr \,e^+ e^-$ and
$\etapr \rightarrow \omega \,e^+ e^-$.

The work is organized as follows. In section \ref{sec:power} we construct the chiral Lagrangian with vector
mesons and the
$\etapr$ meson. After a primer on large-$N_c$ QCD we discuss the naturalness assumption based on the
hadrogenesis conjecture and its
associated mass gap. The complete leading-order Lagrangian is presented. In section \ref{sec:Lagrangian}
we work out the application to the electromagnetic transition form factors of vector to pseudoscalar mesons.

\section{The chiral Lagrangian}
\label{sec:power}

The construction rules of the chiral
$U(3)$ Lagrangian relevant for mesonic systems are readily presented.
For more technical details see for example
\cite{Krause:1990,Ecker:1988te,Ecker:1989yg,Birse:1996hd,HerreraSiklody:1996pm,Kaiser:2000gs}. The basic
building blocks of the chiral Lagrangian are
\begin{eqnarray}
  U_\mu & = & {\textstyle \frac{1}{2}}\,e^{-i\,\frac{\Phi}{2\,f}} \left(
    \partial_\mu \,e^{i\,\frac{\Phi}{f}} \right) e^{-i\,\frac{\Phi}{2\,f}}
  -{\textstyle \frac{i}{2}}\,e^{-i\,\frac{\Phi}{2\,f}} \,r_\mu\, e^{+i\,\frac{\Phi}{2\,f}}  
  \nonumber \\  && {}
  +{\textstyle \frac{i}{2}}\,e^{+i\,\frac{\Phi}{2\,f}} \,l_\mu\, e^{-i\,\frac{\Phi}{2\,f}} \,,
  \qquad  \Phi_{\mu \nu} \;, 
  \label{def-fields}
  \\
  f_{\mu \nu}^{\pm} & = & {\textstyle \frac{1}{2}} \, e^{+i\,\frac{\Phi}{2\,f}}
  \left( \partial_\mu \,l_\nu- \partial_\nu \,l_\nu -i\,[l_\mu ,\,l_\nu]_-\right)
  e^{-i\,\frac{\Phi}{2\,f}}
  \nonumber\\  && {}
  \pm  {\textstyle \frac{1}{2}} \, e^{-i\,\frac{\Phi}{2\,f}}
  \left( \partial_\mu \,r_\nu\,- \partial_\nu \,r_\nu \,-i\,[r_\mu \,,\,r_\nu\,]_-\right)
  e^{+i\,\frac{\Phi}{2\,f}} \,,   \nonumber
\end{eqnarray}
where we include a nonet of  pseudoscalar-meson fields
$\Phi(J^P\!\!=\!0^-)$ and a nonet of vector-meson fields  in the antisymmetric tensor representation
$\Phi_{\mu \nu} (J^P\!\!=\!1^-)$.
The classical source functions $r_\mu$ and $l_\mu$ in (\ref{def-fields}) are
linear combinations of the vector and axial-vector sources of QCD with $r_\mu = v_\mu+a_\mu$ and $l_\mu = v_\mu-a_\mu$.
The merit of the particular field combinations as displayed in (\ref{def-fields}) is their identical transformation property
under $U_L(3)\otimes U_R(3)$ rotations. Electromagnetism is introduced by $v_\mu \to -e \, {\mathcal Q} \, A_\mu$ with the
positron charge $e$ and the three-flavor quark-charge matrix
\begin{eqnarray}
  \label{eq:defQcm}
  {\mathcal Q} = \left(
    \begin{array}{ccc}
\frac23 & 0 & 0\\
0 & -\frac13 & 0 \\
0 & 0 & -\frac13
\end{array}
\right)  \,.
\end{eqnarray}

Since we will assume perfect isospin symmetry it is convenient to decompose
the fields into their isospin multiplets,
\begin{eqnarray}
  \Phi & = & \tau \cdot \pi (140)
  + \alpha^\dagger \cdot  K (494) +  K^\dagger(494) \cdot \alpha
  \nonumber \\ && {}
  + \eta_8\,\lambda_8 
  + \eta_1 \,\sqrt{{\textstyle \frac{2}{3}}}\,\one \,,
  \nonumber\\
  \Phi_{\mu \nu} & = & \tau \cdot \rho_{\mu \nu}(770)
  + \alpha^\dagger \cdot  K_{\mu \nu}(892) +  K_{\mu \nu}^\dagger(892) \cdot \alpha
  \nonumber\\
  && {} + \Big( \sqrt{\textstyle \frac{2}{3}}\,\lambda_0 
  + {\textstyle \frac{1}{\sqrt{3}}} \,\lambda_8  \Big)\, \omega_{\mu \nu}(782) 
  \nonumber\\
  && {} +\Big(
  {\textstyle \frac{1}{\sqrt{3}}}\,\lambda_0-\sqrt{{\textstyle \frac{2}{3}}}\,\lambda_8\Big)\, \phi_{\mu \nu}(1020)  \,,
  \label{def-decomposition}
  \\
  \alpha^\dagger & = & {\textstyle \frac{1}{\sqrt{2}}}\,(\lambda_4+i\,\lambda_5 ,\lambda_6+i\,\lambda_7 )\,,\qquad
  \tau = (\lambda_1,\lambda_2, \lambda_3)\,, \nonumber 
\end{eqnarray}
with for instance the rho triplet $\rho_{\mu \nu} =(\rho_{\mu \nu}^{(1)},\rho_{\mu \nu}^{(2)},\rho_{\mu \nu}^{(3)} )$ and
the kaon doublet $K_{\mu \nu} =(K_{\mu \nu}^{+},\,K^{0}_{\mu \nu})^t$. The matrices
$\lambda_i$ are the standard Gell-Mann generators of the SU(3) algebra for $i=1,...,8 $ and
$\sqrt{{\textstyle \frac{3}{2}}}\,\lambda_0$ is the 3-dimensional unit matrix.
The numbers in brackets recall the approximate masses of the fields in units of MeV. While in (\ref{def-decomposition})
we already identified the physical $\omega$ and $\phi$ meson states, we keep the unphysical singlet state $\eta_1$ and
octet state $\eta_8$ in the decomposition of the field for the pseudoscalar nonet.

Explicit chiral symmetry-breaking effects are included in terms
of scalar and pseudoscalar source fields $\chi_\pm $ proportional to the quark-mass
matrix of QCD
\begin{eqnarray}
\chi_\pm = {\textstyle \frac{1}{2}} \left(
e^{+i\,\frac{\Phi}{2\,f}} \,\chi_0 \,e^{+i\,\frac{\Phi}{2\,f}}
\pm e^{-i\,\frac{\Phi}{2\,f}} \,\chi_0 \,e^{-i\,\frac{\Phi}{2\,f}}
\right) \,,
\label{def-chi}
\end{eqnarray}
where $\chi_0 =2\,B_0\, {\rm diag} (m_u,m_d,m_s) \approx {\rm diag}(m_\pi^2,m_\pi^2,2 \, m_K^2-m_\pi^2)$.
Similarly, we introduce a pseudoscalar flavor-singlet field
\begin{eqnarray}
H = \frac{1}{\sqrt{6}\,f}\,\tr \,\Phi  \,.
\label{def-chiA}
\end{eqnarray}

The chiral Lagrangian consists of all possible interaction
terms, formed with the fields $U_\mu$, $\Phi_{\mu \nu}$ and $\chi_\pm$, $f^\pm_{\mu \nu}$, $H$ and their
respective covariant derivatives. Derivatives of the fields must be included in compliance
with the chiral $U(3)$ symmetry. This leads to the notion of a covariant derivative
$D_\mu$ which is identical for all fields in (\ref{def-fields}) and (\ref{def-chi}).
For example, it acts on the $\chi_\pm $ fields as follows
\begin{eqnarray}
D_\mu \chi_\pm & = & \partial_\mu \, \chi_\pm + [\Gamma_\mu,\chi_\pm]_- \,,
\label{def-covariant-derivative}\\
\Gamma_\mu & =& {\textstyle \frac{1}{2}}\,e^{-i\,\frac{\Phi}{2\,f}} \,
\Big[\partial_\mu -i\,(v_\mu + a_\mu) \Big] \,e^{+i\,\frac{\Phi}{2\,f}}
\nonumber \\ && {}
+{\textstyle \frac{1}{2}}\, e^{+i\,\frac{\Phi}{2\,f}} \,
\Big[\partial_\mu -i\,(v_\mu - a_\mu)\Big] \,e^{-i\,\frac{\Phi}{2\,f}}\,.
\nonumber
\end{eqnarray}
The covariant derivative of the flavor-singlet field $H$ reads
\begin{eqnarray}
D_\mu \, H = \partial_\mu \, H  - {\textstyle{ \sqrt{\frac{2}{3}} }}\,\tr (a_\mu) \,.
\label{def-DchiA}
\end{eqnarray}

The chiral Lagrangian is a powerful tool once it is combined with appropriate
counting rules leading to a systematic approximation strategy.

\subsection{Scale separation in the hadrogenesis conjecture }

The chiral Lagrangian consists of an infinite number of interaction terms with an infinite number of a priori
unknown parameters. Without an estimate for the relevance of each term the Lagrangian is not predictive.
This is where counting rules play a decisive role in an effective field theory. On the one hand, they
should predict the relative importance of an interaction term. On the other hand, a counting scheme should
provide a guide how to approximate a reaction or decay amplitude given a truncated Lagrangian.

The key issue for the success of an effective field theory is a separation of scales. The goal is to identify
soft versus hard scales and expand in ratios of soft over hard scales.
In the following the generic soft scale will be denoted by $Q$. In an effective field theory that is strictly
perturbative the identification of the two scales is straightforward. In chiral perturbation theory the Goldstone
boson masses together
with their four-momenta are the soft scales. The hard scale is the mass of the lightest degree of freedom that is not part of the
Lagrangian. The Lagrangian can be ordered according to the number of derivatives involved in a given interaction
term. The dimensionful parameters scale with inverse powers of the hard scale. This is called the naturalness
assumption, any effective field theory rests on.

The identification of the characteristic scales is more intricate for an effective Lagrangian that is to be used in
non-perturbative applications, like coupled-channel approaches
\cite{Lutz:2001yb,Lutz:2003fm,Danilkin:2010xd,Gasparyan:2010xz,Danilkin:2011fz}.
For instance, a unitarization of the chiral Lagrangian formulated
with Goldstone bosons only, generates scalar mesons dynamically. We consider the masses of the scalar mesons as dynamically generated
scales which should be discriminated from the characteristic hard scale of the Lagrangian. In our case the mass of the lightest degree of
freedom not part of the Lagrangian must be larger than the mass of the scalar mesons. Given an effective Lagrangian, the possible presence of
dynamic scales makes the identification of the characteristic hard scale a difficult problem. Progress may be possible
by a suitable assumption which then needs to be scrutinized. Such an assumption is the hadrogenesis conjecture, which bets on the only
relevance of pseudoscalar and vector meson fields in the chiral Lagrangian
\cite{Lutz:2001dr,Lutz:2003fm,Lutz:2004dz,Lutz:2005ip,Lutz:2007sk}.
Since such a Lagrangian is expected to generate the rest of the
meson spectrum dynamically, the characteristic hard scale is then significantly larger than the vector-meson masses.

The chiral Lagrangian with dynamical vector-meson fields has been studied in some detail
\cite{Bando:1987br,Meissner:1987ge,Ecker:1988te,Ecker:1989yg,Birse:1996hd,Harada:2003jx,Kampf:2009jh,Djukanovic:2010tb}.
In particular, it was shown that the
leading-order interaction of the vector mesons with the Goldstone bosons generates a quite realistic
spectrum of axial-vector resonances \cite{Lutz:2003fm}. A first attempt to provide a systematic
ordering of the interaction terms according to their relevance was proposed by two
of the authors in \cite{Lutz:2008km}. 

Hadrogenesis, if valid, has a specific implication on the
meson spectrum in the large-$N_c$ limit of QCD. Since the effective Lagrangian includes pseudoscalar and vector
mesons only, we expect the large-$N_c$ meson spectrum to exhibit a sizable gap. For instance, in the limit of a
large number of colors, the lowest $J^P=0^+, 1^+, 2^\pm $ states should
have masses much larger than the lowest $J^P=0^-, 1^-$ states.\footnote{To the best of the authors' knowledge, 
the only stringent fact that can be deduced from QCD in
that limit so far is the existence of an infinite tower of sharp states. The masses of those states are unknown at present.
To this extent our assumption of a possible mass gap in that spectrum is not in contradiction to large-$N_c$ QCD.}
The known physical $J^P=0^+, 1^+, 2^\pm $ states would then be generated
dynamically in terms of the pseudoscalar and vector meson fields. The large-$N_c$ states are integrated out and
set the hard scale $\Lambda_{\rm hard}$ of the effective Lagrangian. In turn it may be justified to consider the masses of the
$J^P=0^-, 1^-$ states as soft scales $Q$ and insist on the counting
\begin{align}
D_\mu , \,m_P, \,m_V \sim Q \, ,
\label{def-soft}
\end{align}
with $m_P$ and $m_V$ the typical mass of a pseudoscalar or vector meson, respectively.
The counting (\ref{def-soft}) relies on a sufficiently large
characteristic hard scale
\begin{eqnarray}
\Lambda_{\rm hard} \geq (2-3) \,{\rm GeV} \,,
\label{def-hard}
\end{eqnarray}
which identifies the natural size of dimensionful parameters in the chiral Lagrangian.

\begin{figure}[t]
\centering
\includegraphics[clip=true,width=0.48\textwidth]{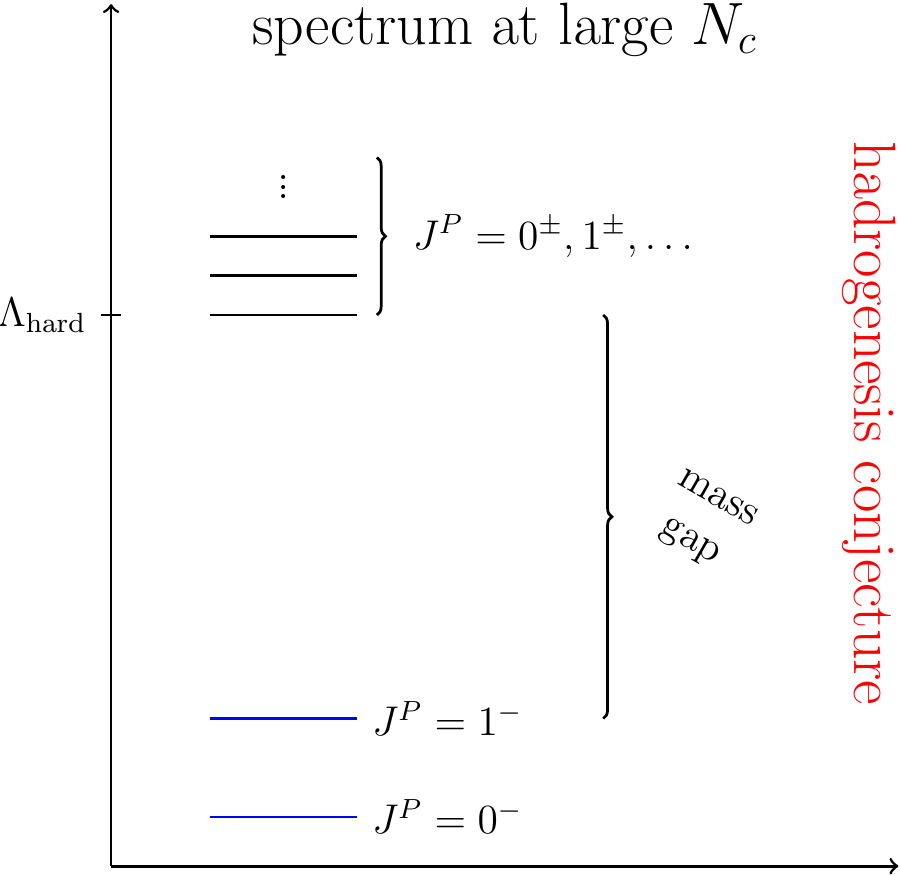}
\caption{The meson spectrum of QCD in the limit of a large number of colors $N_c$ as conjectured in the
hadrogenesis picture.}
 \label{fig:largeNcspectrum}
\end{figure}
In Fig.\ \ref{fig:largeNcspectrum} a possible spectrum of QCD in the large-$N_c$ limit is shown. While the lowest
pseudoscalar and
vector mesons are close to their physical masses, the lowest scalar, axial and tensor states may have significantly larger masses.
The mass gap is characterized by the hard scale
\begin{eqnarray}
\Lambda_{\rm hard} \sim N_c^0  \,,
\label{eq:scalelambdahard}
\end{eqnarray}
which is finite as $N_c$ approaches infinity. Note that such a spectrum does not contradict
Weinberg's sum rules \cite{Weinberg:1967kj}. Our picture would, however,
invalidate a frequent assumption that in the large-$N_c$ limit such sum rules can be saturated by few low-lying states only.

Though at large $N_c$ loop effects are naturally suppressed, it is a nontrivial task to see how this is realized
in terms of the characteristic scales of the system.
In the present work we will focus on the systematic truncation of the chiral Lagrangian. Detailed investigations of loop effects and
renormalization issues will be presented in forthcoming studies.

\subsection{Primer on mesons in large-$N_c$ QCD}

A striking consequence of QCD in the large-$N_c$ limit is  the suppression of $n$-body forces
\cite{'tHooft:1973jz,Witten:1979kh}. A vertex with $n$ meson lines scales with
\begin{eqnarray}
N_c^{1 - n/2} \,.
\label{Nc-counting}
\end{eqnarray}
The simplest application of (\ref{Nc-counting}) is the wave-function normalization, $f_P$ or $f_V$, of a pseudoscalar
or vector meson
\begin{eqnarray}
\frac12 \, \langle 0 | \,\bar q(0) \,\gamma_\mu \,\gamma_5\,\lambda_j \,q(0)\, | P_k(q) \rangle & \equiv &
i \, \delta_{jk} \, q_\mu \, f_P \,,
\label{def-fPV}
\\
\frac12 \, \langle 0 |\, \bar q(0) \,\gamma_\mu \,\lambda_j \, q(0)\, | V_k(q, \lambda) \rangle  & \equiv &
-\delta_{jk}\,m_V\,f_V \,\varepsilon_\mu(q, \lambda)\,, \nonumber
\end{eqnarray}
with the anti-quark field operator $\bar q(x)= (\bar u(x), \bar d(x), \,\bar s(x) )$.
The spin-1 wave function $\varepsilon_\mu(q, \lambda)$ carries momentum $q$, polarization $\lambda$ and mass $m_V$.
It holds
\begin{eqnarray}
f_P \sim \sqrt{N_c}\,, \qquad   f_V \sim \sqrt{N_c}\,, \qquad m_V \sim N_c^0\,.
\label{fPV-scaling}
\end{eqnarray}
In the chiral Lagrangian there will be corresponding tree-level terms
\begin{eqnarray}
 && - f \tr \left\{a_\mu \, \left( (\partial^\mu \Phi)  + \frac{1}{3} \left( \frac{f_H^2}{f^2}-1\right)
 \tr  \,(\partial^\mu \Phi) \right) \right\}
 \nonumber\\
 && {}
+  \frac{1}{2} \, f_V \tr \Big\{ \Phi^{\mu \nu}\,(\partial_\mu \,v_\nu -\partial_\nu \,v_\mu)\Big\} \,,
\label{def-fP-chiral}
\end{eqnarray}
which couple the pseudoscalar and vector meson fields to the external axial and vector source functions $a_\mu$ and $v_\mu$.
A matching of (\ref{def-fP-chiral}) and (\ref{def-fPV}) shows that  $f_P=f $ describes the transition of the
axial-vector current to the
octet Goldstone-boson states and $f_P=f_H$ the corresponding transition to a singlet state. In the strict large-$N_c$ limit \cite{Kaiser:2000gs,Kaiser:1998ds} one
expects $f=f_H$. This illustrates the suppression of terms with multiple flavor traces.

A further application of (\ref{Nc-counting}) with $n=2$ leads to the known scaling of meson masses, as exemplified in (\ref{fPV-scaling}) for the vector mesons and in (\ref{eq:scalelambdahard}) for the masses of the high-energy
large-$N_c$ states. This implies that in general meson masses are not small in the limit of a large number of colors. Yet, there is one important
exception related to the axial anomaly of QCD. The $\eta'$ mass vanishes in the combined large-$N_c$ and chiral
limit \cite{Witten:1979vv}.
Following \cite{Kaiser:2000gs} we consider
\begin{eqnarray}
\label{eq:intro-tau}
  && \tau = -i \int \! d^4x \, \langle 0 \vert  \,T \omega(x) \, \omega(0) \,\vert 0 \rangle\,, \nonumber \\
  && \omega = \frac{g^2}{16\,\pi^2} \, {\rm tr} \left\{G_{\mu\nu} \,\tilde G^{\mu\nu} \right\} \,,
\end{eqnarray}
where $G_{\mu\nu}$ denotes the gluon field strength, $g$ the QCD coupling constant and 'tr' the color trace.
Note that $\tau$ has  mass dimension four and scales with $  \tau \sim N_c^0 $. The non-vanishing of $\tau$ is a result of
the U$_A$(1) anomaly of QCD and introduces the typical mass scale
\begin{eqnarray}
  \label{eq:meta1-tau}
  m_H^2 =  \frac{2 \, N_c \, \tau}{f^2}  \sim N_c^0\,,
  \label{def-mA}
\end{eqnarray}
which is related to the $\etapr$ mass in the large-$N_c$ limit. According to Witten \cite{Witten:1979vv}
and Veneziano \cite{Veneziano:1979ec} the following
relation holds in the chiral limit
\begin{eqnarray}
m^2_{\etapr} \to \frac{N_f}{N_c}\,m_H^2 \sim \frac{1}{N_c}\,,
\label{def-metas}
\end{eqnarray}
with the number of flavors $N_f = 3$. At physical values $N_f = N_c=3$ we may
identify the anomaly scale  $m_H $ with the $\etapr $-meson mass.

While the masses of the pion, kaon and eta mesons are small because of the small up, down and strange quark masses, the $\etapr$ mass is 'small' because
it vanishes at large $N_c$.  Even though the ratio
\begin{eqnarray}
\frac{m_{\eta'}^2}{m_\rho^2} \sim  \frac{1}{N_c} \,,
\end{eqnarray}
vanishes in the large-$N_c$ limit, the empirical masses give a ratio larger than one.
The physical world does not always directly reflect the large-$N_c$ scaling laws
(\ref{eq:scalelambdahard}, \ref{fPV-scaling}, \ref{def-metas}).
This illustrates that the transition from an infinite to the physical number of colors in QCD is intricate and subtle. As we lower $N_c$ down
to its physical value, $N_c =3$, the clear scale separation conjectured in Fig.\ \ref{fig:largeNcspectrum}
is blurred. Scales that are large as compared to
$\Lambda_{\rm hard}$, as, e.g., $f_{P,V}$, may turn semi soft as the number of colors decreases.

\subsection{Naturalness assumption in the  vector-meson sector} \label{subsec:naturalness assumption}

In order to build the chiral Lagrangian with transparent $N_c$-scaling behavior it is useful to introduce a
dimensionless vector-meson field
\begin{eqnarray}
&& V_{\mu \nu} \equiv \frac{1}{f_V}\,\Phi_{\mu \nu} \,,
\label{def-Vmunu}
\end{eqnarray}
which has a seed for the suppression factor (\ref{Nc-counting}). As a consequence, any single-trace contribution
\begin{eqnarray}
f_V^2 \, D^{n_d}(x) \, V^{n_v}(x)
\label{eq:buidblock}
\end{eqnarray}
to the effective Lagrangian has the correct large-$N_c$ scaling behavior. Here $n_d$ denotes the number of
derivatives $D_\mu$ and $n_V$ the number of vector-meson fields $V_{\mu\nu}$.
For clarity the Lorentz indices have not been displayed explicitly in (\ref{eq:buidblock}).

It is not quite immediate how to determine the effective order $Q$ of a given term in the chiral Lagrangian.
For the sake of clarity
we focus first on interaction terms involving vector-meson fields only. The general strategy is to ensure the
suppression (\ref{Nc-counting}) of $n$-point vertices by multiplying the generic structure (\ref{eq:buidblock})
by an appropriate power of $m_V$, which quantifies the soft scale $Q$ according to (\ref{def-soft}). As
further constraints we demand the correct scaling behavior in the large-$N_c$ limit and the correct
dimensionality of each term of the chiral Lagrangian.

We consider a generic vertex of the form
\begin{eqnarray}
  &&D^{n_d}(x) \,V^{n_v}(x)\to
  f_V^{2}\, D^{n_d}(x) \,V^{n_v}(x)
  \nonumber\\[0.5em]
  &&   \times \;
  \left\{
    \begin{array}{ll}
      m_V^{d_v}\,\Lambda_{\rm OZI}^{2-2\,n_t}\,\Lambda_{\rm hard}^{2\,n_t-n_d-d_v} \;\; & 
      {\rm for} \;\; n_d +d_v-2\,n_t\geq 0 \;, 
      \\[1em]
      m_V^{d_v}\,\Lambda_{\rm OZI}^{2-2\,n_t}\,m_{V}^{2\,n_t-n_d-d_v}  \;\;  &  
      {\rm for} \;\; n_d+d_v -2\,n_t < 0\; .
    \end{array}
  \right.  \nonumber \\
  \label{def-generic-V}
\end{eqnarray}
Furthermore we introduce a characteristic dimension
\begin{eqnarray}
&& d_v =  n_v -2\,,
\label{eq:defdv}
\end{eqnarray}
which leads to the suppression of $n$-body forces as expected from large-$N_c$ QCD.
Terms involving $\chi_{\pm}$, which contain the quark-mass matrix,
can be included by $\chi_{\pm} \sim D^2$.
Similarly any external field can be treated with $f_{\mu \nu}^\pm \sim D^2 $.
With (\ref{def-generic-V}) we wish to estimate the natural size of the vertex.
All dimensionful scales are pulled out of the vertex such that the residual dimensionless
coupling constant is expected to be of order one, which is the naturalness assumption.

The scale
\begin{eqnarray}
\Lambda_{\rm OZI} \sim \sqrt{N_c}
\label{def-LambdaV}
\end{eqnarray}
is supposed to quantify OZI violations \cite{Okubo:1963fa,Zweig:1981pd,Iizuka:1966fk}.
The degree of OZI violation is measured by the number of independent
flavor traces $n_t$ in the vertex \cite{Witten:1979kh}.
The general result (\ref{def-generic-V}) makes the large-$N_c$ scaling  of a
generic interaction term explicit. It follows from the scaling behavior of
$f_V$, $\Lambda_{\rm OZI}$, $\Lambda_{\rm hard}$ and $m_V$ as given in
(\ref{eq:scalelambdahard}, \ref{fPV-scaling}, \ref{def-metas}, \ref{def-LambdaV}).

The naturalness assumption (\ref{def-generic-V}) distinguishes between two cases. For $n_d+d_v-2\,n_t < 0$ the 
dimension of the vertex must
be carried by the typical vector-meson mass $m_V$. Positive powers of the hard scale $\Lambda_{\rm hard}$ cannot occur, 
since that scale reflects modes that are integrated out. There is no other scale but $m_V$ to provide the correct dimension 
of the vertex.

What are the consequences of the suggested naturalness estimate (\ref{def-generic-V})? We have to make an assumption on the typical size of
$\Lambda_{\rm OZI} $. If we assume $\Lambda_{\rm OZI}  \sim \Lambda_{\rm hard}$, the effective power of  a  generic interaction
term is $Q^{n_q}$ with 
\begin{eqnarray}
n_q =  \left\{
\begin{array}{ll}
n_d + d_v  \quad \;\; & {\rm for} \;\; n_d +d_v-2\,n_t\geq 0 \;, \\
2\,n_t   \;\;  &  {\rm for} \;\; n_d+d_v -2\,n_t < 0\; .
\end{array}
\right.
\label{def-nq}
\end{eqnarray}

In application of (\ref{def-nq}) we obtain at order $Q^2$ four terms
\begin{eqnarray}
{\mathcal L}_{1} &=& -\frac{1}{4}\,{\tr }\, \Big\{(D^\mu\,\Phi_{\mu \alpha})\,(D_\nu \,\Phi^{\nu \alpha})\Big\}+
\frac{1}{8}\,m_{V}^2\,{\tr } \,\Big\{ \Phi^{\mu \nu}\,\Phi_{\mu \nu}\Big\}
\nonumber\\
&& + \,\frac{1}{8}\,b_D \,{\tr } \,\Big\{ \Phi^{\mu \nu}\,\Phi_{\mu \nu}\,\chi_+\Big\}
+ \frac{1}{2}\,f_V\,{\rm tr} \Big\{\Phi^{\mu\nu}\,f^+_{\mu \nu}\Big\} \,,
 \label{kinetic-term-V}
\end{eqnarray}
relevant for the two-point functions. The order-$Q^2$ terms relevant for higher $n$-point functions will be
presented below after the inclusion of the pseudoscalar degrees of freedom.
The three terms contributing to the free Lagrangian of the vector mesons
are precisely the terms suggested in our previous work \cite{Lutz:2008km}. The parameters
$m_V \simeq 0.764$ GeV and $b_D = 0.92 \pm 0.05$  can be adjusted to recover the empirical vector meson masses. The leading
terms (\ref{kinetic-term-V}) predict degenerate masses of the $\omega$ and $\rho $ mesons. A second mass term involving two flavor traces would
contribute at order $Q^4$. It receives an additional suppression factor $m_V^2/\Lambda_{\rm OZI}^2 \sim Q^2$. Using the empirical mass difference
of the $\omega$ and $\rho $ mesons suggests
\begin{eqnarray}
\Lambda_{\rm OZI} > \Lambda_{\rm hard}\,,
\end{eqnarray}
in support of our working assumption. The parameter
\begin{eqnarray}
f_V = (140 \pm 14)\, {\rm MeV}
\label{res-fV}
\end{eqnarray}
is determined by the electromagnetic decay processes
$\rho_0 \to e^+\,e^-$, $\omega \to e^+\,e^-$ and $\phi \to e^+\,e^-$; see e.g.\ \cite{Lutz:2008km} where a slightly
different notation has been used. The present notation displays the power counting explicitly.

Given $\Lambda_{\rm OZI} > \Lambda_{\rm hard}$ we conclude that in our theory with vector mesons only
the infinite hierarchy of interaction terms starts at order $Q^2$. There is no term counted as order $Q^0$. This is precisely what
is required to render loop effects perturbative in an effective field theory (see e.g.\ \cite{Scherer:2002tk}).

\subsection{Naturalness assumption in the pseudoscalar-meson sector}
\label{subsec:np}

How to proceed and include the pseudoscalar fields? Consider first a generic vertex not involving any vector-meson fields,
\begin{eqnarray}
  &&D^{n_d}(x) \,U^{n_u}(x)\,H^{n_h}(x) \to \;
  f^{2}\, D^{n_d}(x) \,U^{n_u}(x)\,H^{n_h}(x)
  \nonumber\\[0.5em]
  &&   \times \;
  \left\{
    \begin{array}{l}
      \alpha_H^{n_h}\,m_H^{d_h}\,\Lambda_{\rm OZI}^{2-2\,n_t}\,\Lambda_{\rm hard}^{2\,n_t-n_d-n_u-d_h} \\[.7em]
      \qquad \mbox{for} \quad n_u \neq 0  \quad \& \quad  n_d+n_u+d_h-2\,n_t \geq 0  \;,  \\[1.4em]
      \alpha_H^{n_h}\,m_H^{d_h}\,\Lambda_{\rm OZI}^{2-2\,n_t}\,m_H^{2\,n_t-n_d-n_u-d_h}  \\[.7em]
      \qquad \mbox{for} \quad n_u \neq 0 \quad \& \quad n_d+n_u+d_h-2\,n_t < 0   \;,  \\[1.4em]
      \frac{N_c}{2}\,\alpha_H^{n_h}\,m_H^{d_h}\,\Lambda_{\rm hard}^{2\,-n_d-d_h} \\[.7em]
      \qquad \mbox{for} \quad n_u = 0  \quad \& \quad n_d+d_h-2 \geq 0   \;,      \\[1.4em]
      \frac{N_c}{2}\,\alpha_H^{n_h}\, m_H^{2-n_d}   \\[.7em]
      \qquad \mbox{for} \quad n_u = 0  \quad \& \quad n_d+d_h-2 < 0   \,,
    \end{array}
  \right.
  \nonumber\\
  &&
  \label{def-generic-A}
\end{eqnarray}
composed out of $n_d$ derivatives $D_\mu$ and $n_u$ pseudoscalar nonet fields $U_\mu$ and $n_h$ pseudoscalar
flavor-singlet fields $H$. The characteristic dimension is
\begin{eqnarray}
&& d_h =  n_h-2 \,.
\label{eq:defdh}
\end{eqnarray}

Again, the degree of OZI violation
is measured by the number of independent flavor traces $n_t$ in the vertex. The presence of any $H$ field is accompanied by an additional suppression factor $1/N_c$. However, according to Witten \cite{Witten:1979vv}
the suppression factor $1/N_c$ is counter balanced by the
flavor-enhancement factor $2\sqrt{N_f}$. Therefore we identify
\begin{eqnarray}
  \alpha_H = \frac{2}{N_c} \sqrt{N_f} \,, \qquad \alpha_H^{n_h} \sim Q^0   \,, \qquad
  \frac{N_c}{2}\,\alpha_H^{n_h} \sim Q^0   \,, \phantom{m}
\label{counting-alphaA}
\end{eqnarray}
in (\ref{def-generic-A}). While $\alpha_H $ approaches zero at large $N_c$, due to the flavor enhancement we
have $\alpha_H \simeq 1.2$ at $N_f = N_3 = 3$. This suggests to count $\alpha_H \sim Q^0$.

The general result (\ref{def-generic-A}) makes the large-$N_c$ scaling explicit. We emphasize that the correct
large-$N_c$ scaling of the vertex is reproduced by the leading scaling behavior of
$f$, $\alpha_H$, $m_H $ and $\Lambda_{\rm OZI}$, $\Lambda_{\rm hard}$ as given in
(\ref{eq:scalelambdahard}, \ref{fPV-scaling}, \ref{def-mA}, \ref{def-LambdaV}, \ref{counting-alphaA}).

The naturalness assumption (\ref{def-generic-A}) discriminates four different cases. The first two cases with $n_u \neq 0$ are analogous
to the previous consideration (\ref{def-generic-V}) for the vector-meson sector. The presence of any field $H$ is accompanied by the factor $\alpha_H$ as to recover the correct large-$N_c$ scaling behavior of the vertex. For $n_d-2\,n_t < 0$ the dimension of the vertex
must be carried by the typical mass $m_H$. Positive powers of the hard scale $\Lambda_{\rm hard}$ cannot occur, since that scale
reflects modes that are integrated out. There is no other scale but $m_H$ to provide the correct dimension of the vertex. The last
two cases in (\ref{def-generic-A}) with $n_u=0$ are an almost trivial adaptation of the first two cases. The extra factor $N_c/2$ for $n_u=0$ is
required to recover the correct large-$N_c$ scaling of such vertices.

We consider the consequences of the naturalness assumption (\ref{def-generic-A}). First we explore the leading-order
contributions to the two-point functions of the pseudoscalar mesons. There are 6 terms altogether
\begin{eqnarray}
  {\mathcal L}_{2} & = & - f^2\,\tr  \big\{ U_\mu \,U^\mu \big\}
  +\frac{1}{2}\,\big( f_H^2 -f^2\big)\,(D_\mu \,H )\,(D^\mu H)
  \nonumber\\
  && {} - \frac{1}{2}\,f_H^2\,m_H^2 \,H^2
  + \frac{1}{2}\,f^2\,\tr  \big\{\,\chi_+  \big\} 
  \label{eq:L2pseudo}
  \\
  && {} -\frac{1}{\sqrt{6}}\,i\,f^2\,b_H\,H\,\tr \big\{\,\chi_-\big\}
  +  \frac{1}{2}\,f^2\,g_0\tr  \big\{\,\chi_+  \big\} \,H^2,
  \nonumber
\end{eqnarray}
where we recall our normalization conventions
\begin{eqnarray}
  && H = \frac{1}{\sqrt{6}\,f}\,\tr \Phi \equiv \frac{1}{f_H}\,\tilde \eta_1\,,  \quad
  \eta_8 = \frac{1}{2}\,\tr \left\{ \Phi \,\lambda_8 \right\}  \,,
  \nonumber\\
  && \tilde \eta_1 = -\eta \, \sin\theta + \eta' \, \cos\theta\,, \quad \eta_8 =  \eta \, \cos\theta + \eta' \, \sin\theta \,,
  \phantom{mm}
  \label{def-mixing}
\end{eqnarray}
for the singlet and octet fields $\tilde \eta_1$ and $\eta_8$. This leads to the conventional mixing scenario
with one mixing angle; see e.g.\ the discussions in
\cite{Gasser:1984gg,Ametller:1991jv,Kaiser:1998ds,Feldmann:1998vh,Feldmann:1998sh,Escribano:1999nh,Benayoun:1999fv,Benayoun:1999au,Beisert:2001qb,Gerard:2004gx,Escribano:2005qq,Degrande:2009ps}.

After the diagonalization of the singlet and octet fields the parameters $m_H$ and $b_H$ can be dialed
as to recover the empirical $\eta$ and $\etapr$ masses. For the mixing angle one obtains
\begin{eqnarray}
\cos(2\,\theta) = \frac{m_{\etapr}^2 + m_\eta^2 -  \frac23 \, (4 \,m_K^2 - m_\pi^2) }{m_{\etapr}^2 - m_\eta^2}   \,.
\label{eq:dettheta}
\end{eqnarray}
The parameter $f_H$
does not affect the mixing angle. It enters the determination of the $\eta$ and $\etapr$ masses
\begin{eqnarray}
&& m_{\eta'}^2 + m_\eta^2 =  \nonumber \\
&& \qquad m_H^2  +\frac13  \,\frac{f^2}{f_H^2}\, (1-2\,b_H -3\, g_0)\, (2\,m_K^2+m_\pi^2)
\nonumber \\
&& \qquad {} + \frac13 \, (4\, m_K^2 - m_\pi^2)\,,
\nonumber\\
&& 1- b_H = \frac{3}{\sqrt{2}\,4}\,\frac{f_H}{f}\,\frac{m_{\eta'}^2-m_\eta^2}{m_\pi^2-m_K^2} \,\sin (2\,\theta)\,.
\label{eta-masses}
\end{eqnarray}
For given value of $f_H$ the parameters $b_H$ and $m_H$ are determined by the empirical values for
the $\eta$ and $\etapr$ masses.
The effect of the parameter $g_0$ cannot be discriminated from the effect of the parameter $m_H$ and therefore cannot be determined here.
Note for instance that the choice $b_H=0$ and $f_H=f$ will not allow for a precise reproduction of the empirical masses and therefore
a different value of the mixing angle may arise \cite{Gasser:1984gg,Kaiser:1998ds}.

\begin{figure}[t]
\centering
\includegraphics[clip=true,height=0.49\textwidth,angle=-90]{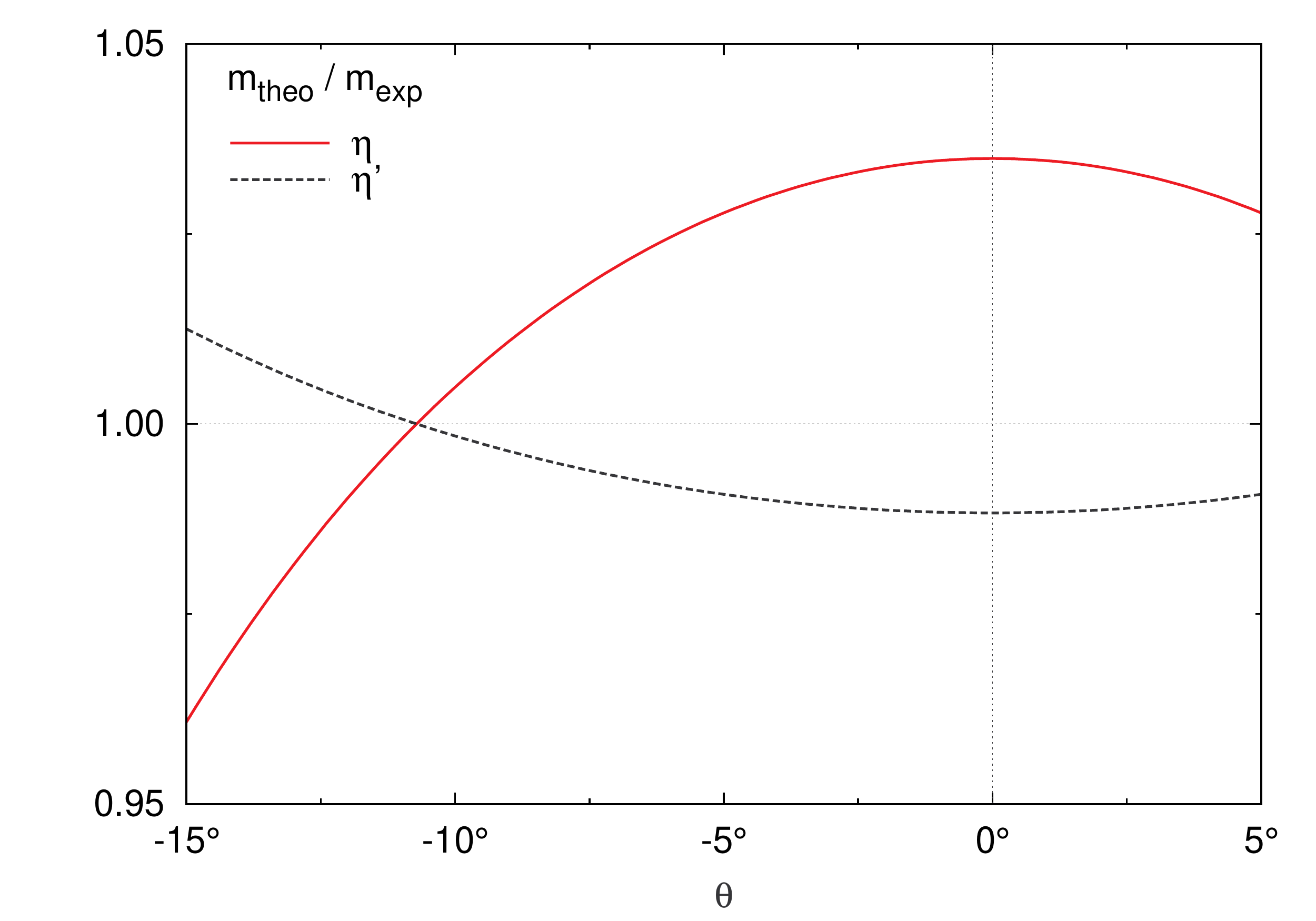}
\caption{The masses of $\eta$ and $\etapr$ as a function of the mixing angle $\theta$. }
 \label{fig:mixing-angle}
\end{figure}
For the empirical masses of $\eta$ and $\etapr$ one would recover from (\ref{eq:dettheta}) the value
$\theta \approx -11^\circ$ \cite{Gasser:1984gg}.
Even though the mixing angle appears to be determined by the empirical $\eta$ and $\etapr$ masses,
it is important to estimate its uncertainty. In Fig.\ \ref{fig:mixing-angle} we show the relative deviation of the
$\eta$ and $\etapr$ masses from their empirical values as a function of the mixing angle $\theta$. Here the squared sum of the
$\eta$ and $\etapr$ masses is kept at its empirical value. The figure shows that the mixing angle has a large uncertainty. If one
accepts an uncertainty in the meson masses at the level of 3$\%$, the mixing angle is compatible with zero degrees.
On the other hand, a $3\%$ deviation of the empirical values from the meson masses obtained in our leading-order approach is surely acceptable. In the following, we treat the mixing angle $\theta$ as a free parameter
which will be determined from radiative vector-meson decays. It will turn out that the resulting
value for $\theta$ is in the acceptable range depicted in Fig.\ \ref{fig:mixing-angle}.

The sector of $\eta$ and $\eta'$ exhibits a rich phenomenology
\cite{Gasser:1984gg,Ametller:1991jv,Kaiser:1998ds,Feldmann:1998vh,Feldmann:1998sh,Escribano:1999nh,Benayoun:1999fv,Benayoun:1999au,Beisert:2001qb,Gerard:2004gx,Escribano:2005qq,Degrande:2009ps}
not all of which can be captured by our leading-order approach of the present work. The kinetic parts of our
Lagrangian \eqref{eq:L2pseudo} are diagonal in the octet-singlet basis. Therefore we have one mixing angle to
obtain the physical states $\eta$ and $\eta'$ from the octet and singlet isoscalar states, see \eqref{def-mixing}.
Expressed in the flavor basis (with an $\eta_q$ made out of up and down quarks and a purely strange $\eta_s$)
the kinetic terms of our Lagrangian would contain non-diagonal parts. This implies two mixing angles in the
flavor basis, but, of course, with a fixed correlation between the two mixing angles (the number of parameters
cannot change by changing the basis). This agrees with the \emph{leading-order} result of large-$N_c$ chiral
perturbation theory \cite{Kaiser:1998ds,Kaiser:2000gs}, but contrasts to phenomenological analyses which suggest
two mixing angles for the octet-singlet basis, but seem to be compatible with one mixing angle in the flavor
basis (see e.g.\ \cite{Escribano:2005qq}).

Such phenomenological analyses are based on data from two-photon decays of pseudoscalars and vector-meson decays
into a photon and a pseudoscalar meson (see next section). In addition, a phenomenological framework allows for
flavor and nonet breaking of the decay constants of the pseudoscalar and the vector mesons. At leading order of
our framework we have one universal decay constant $f_V$ for the whole vector-meson nonet. We also have the same
pseudoscalar decay constant $f$ for the the whole pseudoscalar octet and one separate decay constant $f_H$ for the
$\eta$ singlet. Only at next-to-leading order there will be tree-level and loop structures which allow for separate
vector decay constants for $\rho$, $\omega$ and $\phi$, for $\omega$-$\phi$ mixing, for separate pseudoscalar decay
constants for $\pi$, $K$ and the two $\eta$ states and for two mixing angles in the octet-singlet basis. The situation is
comparable to large-$N_c$ chiral perturbation theory. Also there a two-mixing angle scenario emerges only at
next-to-leading order \cite{Kaiser:1998ds,Kaiser:2000gs} and even then is not in full agreement with the
phenomenological extraction \cite{Escribano:2005qq}.

In addition, the two-photon decays of pseudoscalars are accessible in our framework only at order $Q^4$ and
involve loop diagrams with vector mesons. These decays are the primary source of information in phenomenological
analyses. In contrast, the decays which we can access by a leading-order calculation are the vector-meson decays
into a photon and a pseudoscalar meson. We will find in the next section a rather small mixing angle by studying
these decays. This is significantly different from the phenomenological approaches which extract two mixing angles,
both larger than what we find. Whether a next-to-leading-order calculation in our framework brings the parameters
closer to the present phenomenological results or adds new aspects to the phenomenological analyses,
e.g.\ by loop effects which cause an energy dependence of the mixing angles, remains to be seen.
However, it would not be reasonable to include by hand into a leading-order calculation some effects which are
beyond leading order. Therefore, we stick to a purely leading-order calculation in the present work and use the
obtained leading-order parameters for further predictions.

\subsection{Leading-order effective Lagrangian in the hadrogenesis conjecture}

We are now well prepared to present our naturalness assumption for a vertex involving vector and pseudoscalar fields
\begin{eqnarray}
  && D^{n_d}(x) \,U^{n_u}(x)\,V^{n_v}(x)\,H^{n_h}(x) \nonumber \\[0.7em]
  && \to \;
  f^2\,\alpha_H^{n_h}\,D^{n_d}(x) \,U^{n_u}(x)\,V^{n_v}(x)\,H^{n_h}(x)
  \nonumber\\[0.5em]
  &&   \times \;
  \left\{
    \begin{array}{l}
      m_H^{d_h}\,\Lambda_{\rm OZI}^{2-2\,n_t}\,\Lambda_{\rm hard}^{2\,n_t-n_d-d_h-n_u} \\[.7em]
      \quad \mbox{for} \quad 
      n_u +n_v\neq 0  \quad \& \quad n_d+d_h+n_u-2\,n_t \geq 0  \,, \\[1.4em]
      m_H^{d_h}\,\Lambda_{\rm OZI}^{2-2\,n_t}\,m_H^{2\,n_t-n_d-d_h-n_u}    \\[.7em]
      \quad \mbox{for} \quad 
      n_u +n_v\neq 0 \quad \& \quad n_d+d_h+n_u-2\,n_t < 0  \,,  \\[1.4em]
      \frac{N_c}{2}\,m_H^{d_h}\,\Lambda_{\rm hard}^{2\,-n_d-d_h}  \\[.7em] 
      \quad \mbox{for} \quad
      n_u +n_v= 0  \quad \& \quad n_d+d_h-2 \geq 0  \,,   \\[1.4em]
      \frac{N_c}{2}\,m_H^{2-n_d}   \\[.7em]  
      \quad \mbox{for} \quad 
      n_u+n_v = 0  \quad \& \quad n_d+d_h-2 < 0   \,,
    \end{array}
  \right.  \nonumber \\
  \label{def-generic-VA}
\end{eqnarray}
where we rely on $m_H \sim m_V $ and the generalization of (\ref{eq:defdv}) and (\ref{eq:defdh}) for the
characteristic dimension
\begin{eqnarray}
&& d_h =   n_h + n_v -2\,.
\end{eqnarray}
Again we suppress the Lorentz and flavor structure
in (\ref{def-generic-VA}). Terms involving $\chi_{\pm}$ or $f_{\mu \nu}^\pm $ can be included
by $\chi_{\pm}\sim f_{\mu \nu}^\pm  \sim D^2$.

The interaction terms given in (\ref{def-generic-VA}) scale with $Q$ as
\begin{eqnarray}
Q^{n_q} \quad {\rm with} \;\;  n_q =  \left\{
\begin{array}{l}
{\rm max}(n_d + d_h \,,2) \quad \mbox{for} \; n_u = n_v = 0 \,,  \\
{\rm max}(n_d+n_u+d_h \, ,2\,n_t) \quad \mbox{else.} 
\end{array}
\right.   \nonumber \\
\label{def-nq2}
\end{eqnarray}
We recall that $n_d$ counts the number of derivatives, $n_u$ the number of $U_\mu$ fields, $n_V$ the number
of vector-meson fields $\Phi_{\mu\nu}$, $n_H$ the number of explicit pseudoscalar flavor-singlet fields $H$,
and $n_t$ the number of separate flavor traces in the sector of $U_\mu$ and $\Phi_{\mu\nu}$ fields. Obviously,
$n_t \ge 1$ if $n_u \neq 0$ and/or $n_v \neq 0$. Consequently there are no terms with $n_q < 2$.

As an application of the naturalness assumption (\ref{def-generic-VA}) we work out the leading-order three-point vertices. Technically one can use the developments of \cite{Fearing:1994ga,Bijnens:1999sh,Ebertshauser:2001nj}
to construct a complete set of terms.
Altogether we find 8 terms of order $Q^2$,
{\allowdisplaybreaks\begin{eqnarray}
  {\mathcal L}_3 & = & \frac{i}{2}\,f_V\, h_{P}\,{\rm tr}\,\Big\{U_\mu\,\Phi^{\mu\nu}\,U_\nu\Big\}
  \nonumber \\
  && {}
  +\ \frac{i}{8} \,h_A \,\varepsilon^{\mu\nu\alpha\beta}\,
  {\rm tr} \,\Big\{ \big[\Phi_{\mu \nu},\,(D^\tau \Phi_{\tau \alpha})\big]_+ \,U_\beta\Big\}
  \nonumber \\
  && {}
  - \frac{i}{4} \,\frac{m^2_V}{f_V} \, h_V \,{\rm tr}\,\Big\{
  \Phi_{\mu \tau}\,\Phi^{\mu \nu}\,\Phi^{\tau}_{\;\;\, \nu} \Big\}
  \nonumber \\
  && {}
  - \frac{1}{4} \,m_V^2\, h_H \, \varepsilon^{\mu\nu\alpha\beta}\, {\rm tr} \,\Big\{\Phi_{\mu\nu}\,\Phi_{\alpha \beta}\Big\} \, H
  \nonumber \\
  && {} + \frac{i}{8} \, h_O \, \varepsilon^{\mu \nu \alpha \beta}\,
  \tr \Big\{ \big[ (D_\alpha \Phi_{\mu \nu}), \, \Phi_{\tau \beta}\big]_+ \, U^\tau\Big\}
  \nonumber \\
  && {}
  +  \frac{i}{4} \,b_A \,\varepsilon^{\mu\nu\alpha\beta} \,
  {\rm tr} \,\Big\{ \Phi_{\mu\nu}\, \chi_-\, \Phi_{\alpha \beta} \Big\}
  \nonumber \\
  && {} + \frac{ i}{2} \,e_M\,  {\rm tr}\,\Big\{\Phi^{\alpha}_{ \;\; \tau}\,f^+_{\alpha \beta} \,\Phi^{\tau \beta}\Big\}
  \nonumber \\
  && {}
  -\frac{1}{2} \, f_V\,e_H \, \varepsilon^{\mu\nu\alpha\beta}\, {\rm tr} \,\Big\{\Phi_{\mu\nu}\,f^+_{\alpha \beta}\Big\} \, H \,,
  \label{def-L3}
\end{eqnarray}}
where five terms have been suggested already in our previous work \cite{Lutz:2008km}.
Two out of the three new terms involve the $H$ field.
The remaining new term is proportional to $h_O$ and contributes for processes where at least one vector meson
is off-shell. In order
to have a more transparent realization of our counting rules we have modified our previous notation.
Given the notation of (\ref{def-L3}) the counting is straightforwardly implied by $m_V^2 \sim Q^2$
together with $D_\mu, U_\mu \sim Q$ and $ \chi_\pm , f_{\mu \nu}^\pm \sim Q^2$.

We turn to the four-point vertices. A complete list of terms at order $Q^2$ is
{\allowdisplaybreaks\begin{eqnarray}
    {\mathcal L}_4 & = &
    \frac{1}{8} \, g_1 \,
    {\tr } \,\Big\{\big[ \Phi_{\mu \nu }\,,U_\alpha \big]_+ \, \big[U^\alpha, \Phi^{\mu \nu} \big]_+\Big\}
    \nonumber\\[0.5em]
    && {}       
    +\frac{1}{8} \, g_2 \,
    {\tr } \,\Big\{\big[ \Phi_{\mu \nu }\,,U_\alpha \big]_- \, \big[U^\alpha, \Phi^{\mu \nu}\big]_- \Big\}
    \nonumber\\[0.5em]
    && {} 
    + \frac{1}{8} \, g_3 \,
    {\tr } \,\Big\{\big[\,U_\mu\,,U^\nu \big]_+ \, \big[\Phi_{\nu \tau}\,, \Phi^{\mu \tau}\big]_+  \Big\}
    \nonumber\\[0.5em]
    && {} 
    +\frac{1}{8} \, g_4 \,
    {\tr } \,\Big\{\big[\,U_\mu\,,U^\nu \big]_- \, \big[\Phi_{\nu \tau} \,, \Phi^{\mu \tau}\big]_- \Big\}
    \nonumber\\[0.5em]
    && {} + \frac{1}{8} \, g_5 \,
    {\tr } \,\Big\{\big[\Phi^{\mu \tau} , U_\mu\big]_- \, \big[\Phi_{\nu \tau} \,, U^\nu\big]_- \Big\}
    \nonumber\\[0.5em]
    && {} + \frac18 \, \frac{m_V^2}{f_V^2} \, g_6 \,
    {\tr } \,\Big\{\big[\Phi_{\mu \nu} \,, \Phi_{\alpha \beta} \big]_+ \,
    \big[\Phi^{\alpha \beta} , \Phi^{\mu \nu}\big]_+ \Big\}
    \nonumber\\[0.5em]
    && {}
    + \frac18 \, \frac{m_V^2}{f_V^2} \, g_7 \,
    {\tr } \,\Big\{\big[\Phi_{\alpha \beta} \,, \Phi_{\mu \nu} \big]_- \,
    \big[\Phi^{\alpha \beta} ,\Phi^{\mu \nu}\big]_- \Big\}
    \nonumber\\[0.5em]
    && {} + \frac18 \, \frac{m_V^2}{f_V^2} \, g_8 \,
    {\tr } \,\Big\{ \big[\Phi^{\mu \nu} ,\Phi_{\mu \beta} \big]_+ \,
    \big[\Phi_{\alpha \nu} \,, \Phi^{\alpha \beta}\big]_+ \Big\}
    \nonumber\\[0.5em]
    && {}
    + \frac18 \, \frac{m_V^2}{f_V^2} \, g_9 \,
    {\tr } \,\Big\{ \big[\Phi^{\mu \nu} ,\Phi_{\mu \beta} \big]_- \,
    \big[\Phi_{\alpha \nu} \,, \Phi^{\alpha \beta}\big]_- \Big\}
    \nonumber\\[0.5em]
    && {} + \frac{i}{4} \, f \, g_{10} \, \varepsilon^{\mu \nu \alpha \beta} \,
    \tr \,\Big\{\Phi_{\mu \nu}\,U_\alpha \,U_\beta \Big\} \,H
    \nonumber\\[0.5em]
    && {}
    + \frac{i}{4} \, \frac{m_V^2}{f_V} \, g_{11} \, \varepsilon^{\mu \nu \alpha \beta}  \,
    \tr \, \Big\{ \Phi_{\mu \nu}\,\Phi_{\alpha \tau}\,\Phi^{\tau}_{\phantom{\tau} \beta} \Big\} \,H
    \nonumber\\[0.5em]
    && {} + \frac{1}{4} \, f^2 \, g_{12} \,
    {\tr } \,\Big\{U_\mu\,U^\mu \Big\}\,H^2
    \nonumber\\[0.5em]
    && {}
    +\frac{1}{4}\,m_V^2\,g_{13}\,{\tr } \,\Big\{\Phi_{\mu \nu}\,\Phi^{\mu \nu} \Big\}\,H^2
    + f_H^2 \, g_{14} \, H^4 \,,
    \label{def-L4}
  \end{eqnarray}}
where the first two terms have been constructed already in our previous work \cite{Lutz:2008km}.

Note that the terms (\ref{kinetic-term-V}, \ref{eq:L2pseudo}, \ref{def-L3}, \ref{def-L4})
form together the {\em complete} leading-order Lagrangian of our approach. It is of order $Q^2$, which allows for
a perturbative treatment of loop contributions. One-loop diagrams start to contribute at order $Q^4$.
This is in contrast to a strict large-$N_c$ counting \cite{Kaiser:2000gs} where inverse powers of $f$
are treated as a soft scale. In such a scheme loops contribute only at next-to-next-to-leading order.
In our approach the large-$N_c$ suppression of $n$-point vertices (\ref{Nc-counting}) has been mapped on
the soft scale $m_V \sim m_H$ in the naturalness assumption (\ref{def-generic-VA}).
Inverse powers of $f$, $f_V$ or $f_H$ are not treated as soft in
the final power counting (\ref{def-nq2}).

We emphasize that, on the one hand, multiple-trace terms do not
contribute at the same order as single-trace terms for the $Q^2$ counter terms. This reflects the phenomenological OZI rule.
On the other hand, multiple-trace terms do contribute at subleading order $Q^4$, as is required to renormalize the effective Lagrangian in
perturbation theory. According to the naturalness assumption (\ref{def-generic-VA}) all terms involving four $U_\mu$ fields
are of order $Q^4$, like they are in the Gasser-Leutwyler Lagrangian. In fact all the 10 subleading-order terms of the conventional chiral
Lagrangian turn relevant at the same order $Q^4$ in our scheme. This includes all terms with two flavor traces and is in line with the finding \cite{Gasser:1983yg,Ecker:1988te} that the size of the symmetry-conserving counter terms
can be estimated by tree-level exchange processes of
the vector mesons, if interpolated by antisymmetric tensor fields. Once vector-meson fields are included explicitly in the chiral Lagrangian the
residual counter terms have a much reduced size as compared to the estimates of
Gasser and Leutwyler \cite{Gasser:1983yg,Gasser:1984gg}. Indeed in a
recent coupled-channel computation based on our Lagrangian truncated at the order $Q^2$, the empirical phase shifts for pion  and kaon scattering
were reproduced quite accurately without the need of tuning any free parameters \cite{Danilkin:2011fz}.

We stress again that a strictly perturbative application of our Lagrangian would not always lead to meaningful
results. The dynamical generation of states requires a coupled-channel framework where rescattering
effects are resummed to all orders. Only for the two-particle irreducible scattering kernel and not for the
scattering amplitude a perturbative expansion can make sense. Yet there is a regime where one can expect that the
rescattering effects are less important than for the scattering region, namely the low-energy regime of particle
decays. Therefore, the strategy is to determine as many parameters as possible by matching perturbative
calculations of
decay amplitudes to experimental data. In addition, one can obtain predictions for other
decay processes which further test the proposed power counting scheme.
Of course, one always has to check that rescattering effects are really subdominant for the considered
decays  \cite{Leupold:2008bp,Terschluesen:2010ik}. This strategy provides input parameters for the coupled-channel
calculations relevant for the scattering regime \cite{Danilkin:2011fz}.
Explicit next-to-leading-order calculations including loops are postponed to
future works. At present we focus on phenomenological consequences of our leading-order Lagrangian.

\newpage

\section{Transition form factors for radiative decays}
\label{sec:Lagrangian}

The transition matrix element for the decay of a vector meson $V$ into a pseudoscalar meson $P$ and a dilepton $l^+ l^-$ can be parametrized as \cite{Landsberg:1986fd}
\begin{eqnarray}
  \mathcal{M}_{V \rightarrow P \,l^+ l^- } & = & -e^2 \, f_{VP}(q^2) \, \eps^{\,\mu\nu\alpha\beta} \,q_\mu \,k_\nu \, \eps_\alpha (q+k, \lambda)\,
\nonumber \\ && \times
  \frac{1}{q^2} \, \bar{u}(q_1, \lambda_1) \,\gamma_\beta \,v(q_2, \lambda_2)\, ,
\label{eq:matrix element}
\end{eqnarray}
with a form factor $f_{VP}(q^2)$. Here, $q$ and $k$ are the four-momenta of the virtual photon and the
pseudoscalar meson $P$, respectively. The wave function of the vector meson is $\eps_\alpha(q+k, \lambda)$ and
$ u(q_1,\lambda_1), v(q_2, \lambda_2)$ denote the wave functions of the two leptons with
their respective four-momenta $q_{1,2}$ .

The double-differential decay rate of the decay of a vector meson $V$ with mass $M_V$ into a pseudoscalar meson
$P$ with mass $M_P$ and a dilepton $l^+ l^-$
is given as \cite{Nakamura:2010zzi}
\begin{eqnarray}
&&\frac{\te{d}^2 \Gamma_{V \rightarrow P \,l^+ l^-}}{\te{d}q^2 \, \te{d}m_{l^+ P}^2} =
 \frac{e^4}{(2\pi)^3} \, \frac{1}{32 \, M_V^3} \,
\left| f_{VP}(q^2) \right| \, \frac{P}{q^4}\, , \nonumber \\[0.5em]
&& m_{l^+ P}^2 = (q_2 + k)^2\,,
  \label{eq:dd width}
\\[0.5em]
&&  P = - \frac{1}{3} \, \eps^{\,\mu\nu\alpha\beta} \,
q_\mu \,k_\nu \, \eps_{\bar{\mu}\bar{\nu}\bar{\alpha}\bar{\beta}} \,
 q^{\,\bar{\mu}} \,k^{\bar{\nu}} \, g_\alpha^{\phantom{\alpha}\bar\alpha} \,
\nonumber\\
&& \quad \times \,
 \sum_{\lambda_1, \lambda_2} \bar{u}(q_1, \lambda_1) \, \gamma_\beta \, v(q_2, \lambda_2) \,
 \bar{v}(q_2, \lambda_2) \,
 \gamma^{\,\bar{\beta}} \, u(q_1, \lambda_1) \,.  \nonumber
\end{eqnarray}
The single-differential decay width \cite{Landsberg:1986fd}
\begin{eqnarray}
  && \frac{\te{d} \Gamma_{V \rightarrow P \, l^+ l^-}}{\te{d}q^2 \, \Gamma_{V \rightarrow P \gamma}}
  = \left| F_{VP}(q^2) \right|^2 \, 
  \sqrt{1- \frac{4\, m_l^2}{q^2}} \, \left( 1+ \frac{2 \,m_l^2}{q^2} \right) \, \frac{1}{q^2}  \nn \\
  && \times  \frac{e^2}{3 \, (2\pi)^2} \,\left[ \left( 1+ \frac{q^2}{M_V^2 -M_P^2} \right)^2
    - \frac{4 \, M_V^2 \, q^2}{\left(M_V^2-M_P^2\right)^2} \right]^{3/2}  \,,
  \nonumber \\[0.3em]
  &&  F_{VP}(q^2) = \frac{f_{VP}(q^2)}{f_{VP}(q^2=0)} \,, \nonumber \\[0.3em]
  &&  \Gamma_{V \rightarrow P\, \gamma} = \frac{\left(M_V^2-M_P^2\right)^3 }{96 \,\pi \, M_V^3} \, \big| e\,f_{VP}(0) \big|^2 \,,
  \label{eq:sd width}
\end{eqnarray}
is obtained by integrating \eqref{eq:dd width} in the kinematically allowed region.
It includes the lepton mass, $m_l$, and is normalized
to the partial decay width for the decay into a real photon.
The normalized form factor $F_{VP}(q^2)$
is equal to $1$ at the photon point $q^2 = 0$. The result (\ref{eq:sd width}) is applicable for the case with
$M_V > M_P$. In the opposite case with $M_P > M_V$ the decay $P \to V \,l^+l^-$ may be possible and
an expression analogous to (\ref{eq:sd width}) is obtained by the exchange $M_V^2 \leftrightarrow M^2_P$
and replacing $\Gamma_{V\rightarrow P \,\gamma}$ by
\begin{align}
 \Gamma_{P \rightarrow V \gamma} = \frac{\left(M_P^2-M_V^2\right)^3 }{32 \pi \, M_P^3} \,\Big| e\,f_{VP}(0) \Big|^2 \,,
 \label{eq:width gamma pseudo}
\end{align}
which reflects the different spins of vector and pseudoscalar mesons. The form factor $f_{VP}(q^2)$ is universally defined by
(\ref{eq:matrix element}).

In this work we consider five decay processes characterized by their form factors
\begin{eqnarray}
  && f_{\omega \,\pi^0}(q^2) = 
  \frac{1}{2\, m_\omega } \, \frac{f_V}{f} \left\{ \left( m_\omega^2 + q^2 \right) h_A - 8 \, m_\pi^2 \, b_A \right\} \,
  S_{\rho}(q^2)\, ,
  \nonumber\\
  && f_{\omega \,\eta_8}(q^2) = 
  \frac{1}{6 \, \sqrt{3} \,m_\omega } \, \frac{f_V}{f} \, \left\{  \left( m_\omega^2 + q^2 \right) h_A
    - 8 \, m_\pi^2 \, b_A  \right\}  \nonumber \\ && 
  \qquad \qquad \times \; S_{\omega}(q^2) \,,
  \nonumber\\
  && f_{\omega \,\eta_1}(q^2) = 
  \frac{1}{3 \, \sqrt{6} \, m_\omega } \, \frac{f_V}{f} \, S_{\omega}(q^2) \nonumber \\
  && \qquad \qquad \; \times \; \left\{
    \left( m_\omega^2 + q^2 \right) h_A - 8 \, m_\pi^2 \,b_A - 4 \,\sqrt{6} \, m_V^2 \,h_H   \right\}  
  \nonumber \\ && \qquad \qquad {} 
  -\frac{2}{3  \, m_\omega}\, \frac{f_V}{f} \, e_H \,,
  \nonumber\\
  && f_{\phi\,\eta_8}(q^2) = 
  \frac{2}{3 \, \sqrt{6} \, m_\phi } \, \frac{f_V}{f} \, S_{\phi}(q^2) \nonumber \\
  && \qquad \qquad \times \; \left\{ \left(m_\phi^2 + q^2 \right) h_A
  - 8 \left(2 m_K^2 -m_\pi^2 \right) b_A  \right\} \,,
  \nonumber\\
  && f_{\phi\,\eta_1}(q^2) = - \,\frac{1}{3 \, \sqrt{3} \, m_\phi } \, \frac{f_V}{f} \, S_{\phi}(q^2) 
  \nonumber \\ && 
  \; \times \; 
  \left\{ \left(m_\phi^2 + q^2 \right) h_A - 8 \left(2 m_K^2 -m_\pi^2 \right) b_A
    - 4 \,\sqrt{6} \, m_V^2\, h_H   \right\} 
  \nonumber \\ && {}
  + \frac{2 \, \sqrt{2}}{3 \, m_\phi} \, \frac{f_V}{f}\, e_H \, ,
  \label{eq:f-all}
\end{eqnarray}
with
\begin{eqnarray}
  &&  f_{\omega \,\eta} = \cos\theta \, f_{\omega \, \eta_8} \, - \, \frac{f}{f_H} \, \sin\theta \, f_{\omega \,\eta_1}\,, 
  \nonumber \\ &&
  f_{\omega  \,\etapr} = \sin\theta \, f_{\omega \,\eta_8} \, + \, \frac{f}{f_H} \, \cos\theta \, f_{\omega \,\eta_1}\,,
  \nonumber\\
  &&  f_{\phi \,\eta} = \cos\theta \, f_{\phi \,\eta_8} \, - \, \frac{f}{f_H} \, \sin\theta \, f_{\phi \,\eta_1}\,, 
  \nonumber \\ &&
  f_{\phi \, \etapr} = \sin\theta \, f_{\phi \,\eta_8} \, + \, \frac{f}{f_H} \, \cos\theta \, f_{\phi \,\eta_1}\,.
  \nonumber\\
  &&   \label{eq:f-mixing}
\end{eqnarray}
The $\rho$-meson propagator \cite{Leupold:2008bp} is
\begin{align}
 && S_{\hspace{-0.1em}\rho}(q^2) = \frac{1}{q^2-m_\rho^2 + i \,\sqrt{q^2}\, \Gamma_\rho(q^2)}\,, \nonumber \\[0.5em]
 && \Gamma_\rho(q^2) = \left[ \frac{q^2-4\,m_\pi^2}{m_\rho^2-4\,m_\pi^2} \right]^{3/2} \frac{m_\rho^2}{q^2}\,\Gamma_0 \,,
\end{align}
with an energy-dependent width. The on-shell width, $\Gamma_0$, of the $\rho$ meson
is approximately equal to $150 \, \te{MeV}$ \cite{Nakamura:2010zzi}.
The $\omega$ and $\phi$ meson propagators $S_{\omega}(q^2) $,
$S_{\phi}(q^2) $ in (\ref{eq:f-all}) are taken in the constant-width approximation.

The empirical decay widths \cite{Nakamura:2010zzi}
of the processes $\omega \to \pi^0 \,\gamma$, $\omega \to \eta \,\gamma$,
$\etapr \to \omega \, \gamma$, $\phi \to \eta \,\gamma$, and $\phi \to \etapr \,\gamma$
imply the matrix elements
{\allowdisplaybreaks\begin{eqnarray}
\big|f_{\omega \,\pi^0} \big| &=& (2.302 \pm 0.040) \,{\rm GeV}^{-1}  \,, 
\nonumber\\
\big|f_{\omega \,\eta}\big| &=& (0.449 \pm 0.020) \,{\rm GeV}^{-1} \,,
\nonumber\\
\big|f_{\omega \,\etapr}\big| &=& (0.431 \pm 0.020) \,{\rm GeV}^{-1}\,, 
\nonumber\\
\big| f_{\,\phi \,\eta}\big| &=& (0.694 \pm 0.007) \,{\rm GeV}^{-1}  \,,
\nonumber\\
\big|f_{\,\phi \,\etapr}\big| &=& (0.723 \pm 0.013) \,{\rm GeV}^{-1} \,,
 \label{empirical-fs}
\end{eqnarray}}
evaluated at the photon point with $q^2 =0$.

We adjust the five parameters $h_A$, $b_A$, $h_H$, $e_H$ and $\theta $ to the five empirical decay amplitudes
(\ref{empirical-fs}) for a given choice of the
parameter $f_H$. Strictly speaking we determine in this way the parameter combinations
$f_V \, h_A$, $f_V \, b_A$, \ldots as can be seen from (\ref{eq:f-all}).
For $f_V$ we use the central value given in (\ref{res-fV}).
The central values of the empirical amplitudes (\ref{empirical-fs}) can always be reproduced exactly. Since the phases of the
decay amplitudes are not determined by the five decay widths there exist $2^5$ distinct solutions reflecting the various phase choices. Using the
phase convention of our previous work \cite{Lutz:2008km}
with $h_A>0$ there remain two physical relevant scenarios only. We reject solutions with unnatural large
parameters that rely on significant cancellation effects. In the two relevant scenarios we find the parameters
\begin{eqnarray}
&& h_A = 2.33 \pm 0.03 \,, \quad b_A =0.16 \pm 0.01 \,, \nonumber \\
&& \theta =  \pm \,(2.0 \pm 1.1) \,,
\label{res-hA}
\end{eqnarray}
independent on the choice of $f_H$. This follows from the independence of the three octet
amplitudes $f_{\omega \,\pi^0}, f_{\omega \,\eta_8}$ and $f_{\phi\,\eta_8}$ on the parameter $f_H$. While the parameter $h_A$ is mainly determined
by the $\omega \to \pi^0 \,\gamma$ decay, the parameter $b_A$ is required to reproduce the $\phi\to \eta \,\gamma$ decay. The decay
$\omega \to \eta \,\gamma$ then requests our small mixing angle in (\ref{res-hA}). We observe that the mixing angle comes at
a value that is quite compatible with the empirical $\eta$ and  $\etapr$ masses.
According to Fig.\ \ref{fig:mixing-angle} the deviation from the empirical masses is less than $3\%$
for either sign of the mixing angle. The theoretical error bars in (\ref{res-hA}) are determined such that
the $\chi^2/N$ of the five considered decays remains smaller than 1 if any single parameter is varied within the suggested error interval.

We turn to the remaining two parameters $h_H $ and $e_H$, which are essentially determined by the two decay
processes involving the $\etapr$ meson.  We find
\begin{eqnarray}
&& e_H = -0.20 \mp ( 0.70 \pm 0.02 ) \,\frac{f_H}{f}  \,, \nonumber \\ 
&& h_H = 0.14 \mp (0.19 \pm 0.01)\,\frac{f_H}{f}  \,,
\label{res-eH}
\end{eqnarray}
where the signs in (\ref{res-hA}) and (\ref{res-eH}) are correlated.  This implies for instance that for the
negative mixing-angle scenario the parameters
$e_H$ and $h_H$ are positive for $f=f_H$. We note that we rejected a pair of solutions where the coefficients in front of the $f_H$ terms
in (\ref{res-eH}) are about three times as large as in our natural solutions.

Since so far there is no means in our scheme to determine the
parameter $f_H$, in the following we assume a natural range with
\begin{eqnarray}
f \leq f_H \leq \sqrt{2}\,f \,,
\label{f0-range}
\end{eqnarray}
where we recall that $f_H  \to f$ in the limit of a large number of colors. For the upper limit used in
(\ref{f0-range}) one finds $f_H \approx f_V$, cf.\ (\ref{res-fV}). As we will see below, electromagnetic
transition form factors of vector to pseudoscalar mesons could provide a tool to further constrain the value of
$f_H$.

For the parameter determination we did not include the $\pi \, \gamma$ decay of the $\rho$ meson.
The decay amplitudes
of the $\omega $ and $\rho$ decays transform into each other upon the interchange of the $\rho$ and $\omega$ mesons.
The result is compatible with the available data.
Since the data involving the $\omega$ have smaller error bars there is no point
to use  the $\rho$-meson decays in the parameter determination.

We have not used  $K^* \to K \, \gamma$ decays \cite{Nakamura:2010zzi} to determine our parameter set.
The large width of the $K^*$  makes it very difficult to extract a model-independent branching ratio.
Given the values for $f_V$, $h_A$ and $b_A$ in (\ref{res-fV}, \ref{res-hA}) the PDG value for the
$K^*_0 \to K_0 \gamma$ decay is reproduced within our theoretical uncertainties. However, the empirical ratio
\begin{eqnarray}
  \left. \frac{\Gamma_{K^*_\pm \to K_\pm \gamma}}{\Gamma_{K^*_0 \to K_0 \gamma}} \right\vert_{\rm exp} \simeq 0.43 \,,
  && \quad \left. \frac{\Gamma_{K^*_\pm \to K_\pm \gamma}}{\Gamma_{K^*_0 \to K_0 \gamma}} \right\vert_{\rm theo} \simeq 0.79 \,,
  \phantom{m} 
  \label{res-ratio}
\end{eqnarray}
cannot be reproduced in this leading-order computation. The theoretical ratio in (\ref{res-ratio}) cannot be changed by any variation
of our parameters (see also the discussion in \cite{Benayoun:1999fv,Benayoun:1999au}).

As already announced at the end of subsection \ref{subsec:np} our leading-order mixing angle given in \eqref{res-hA}
is significantly smaller than what is extracted in phenomenological
analyses \cite{Feldmann:1998vh,Feldmann:1998sh,Escribano:2005qq}. In appendix \ref{sec:appendix} we show that the
use of a larger mixing angle in our leading-order treatment would lead to a much worse description of the real-photon
decay amplitudes \eqref{empirical-fs}. Therefore we stick to our rather small mixing angle for the following
form factor predictions.

\subsection{Decay $\omega \rightarrow \pi^0 l^+ l^-$}

\begin{figure}[h!]
\center
\includegraphics[clip=true,width=0.3\textwidth]{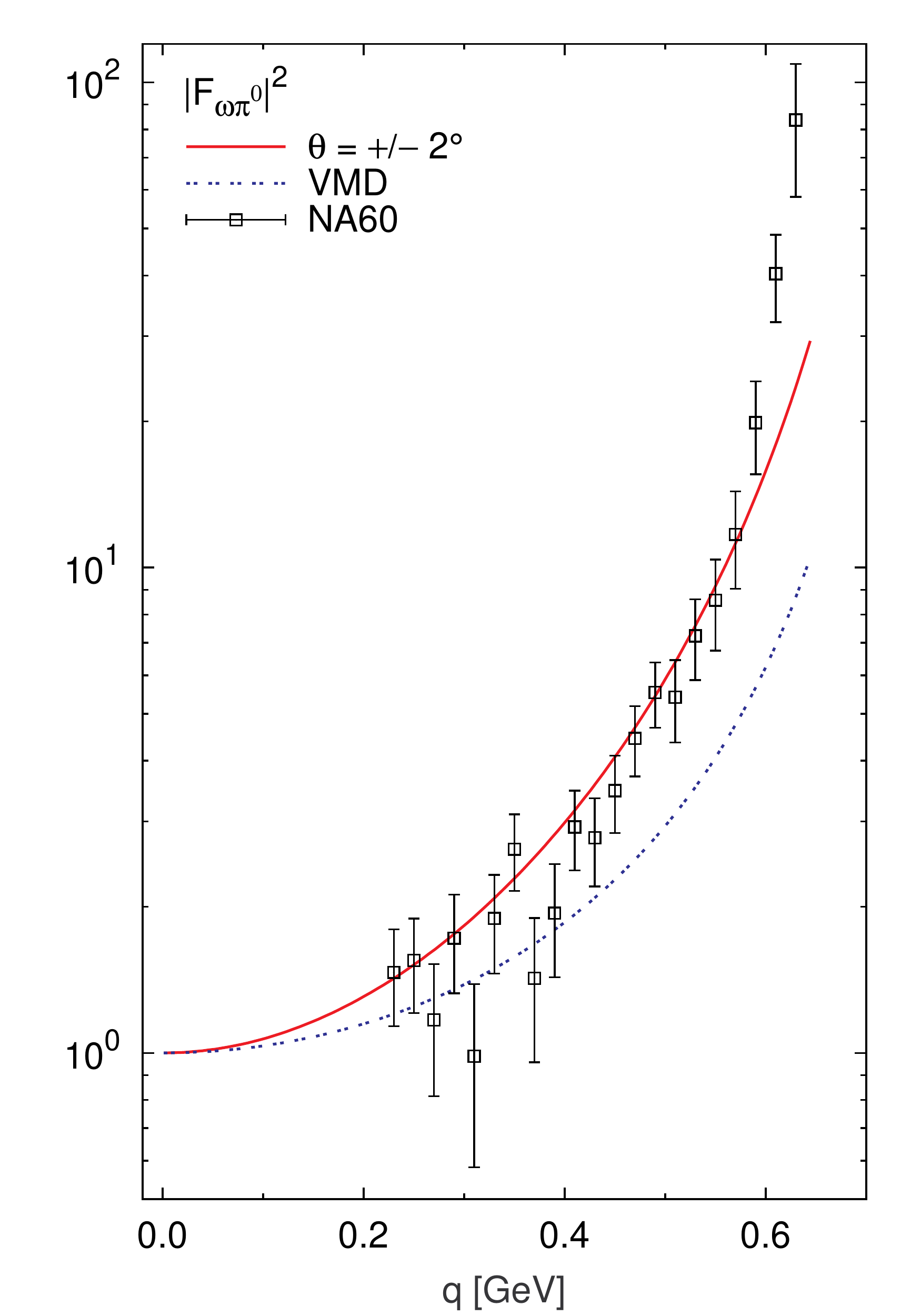} \\
\includegraphics[clip=true,height=0.46\textwidth, angle = -90]{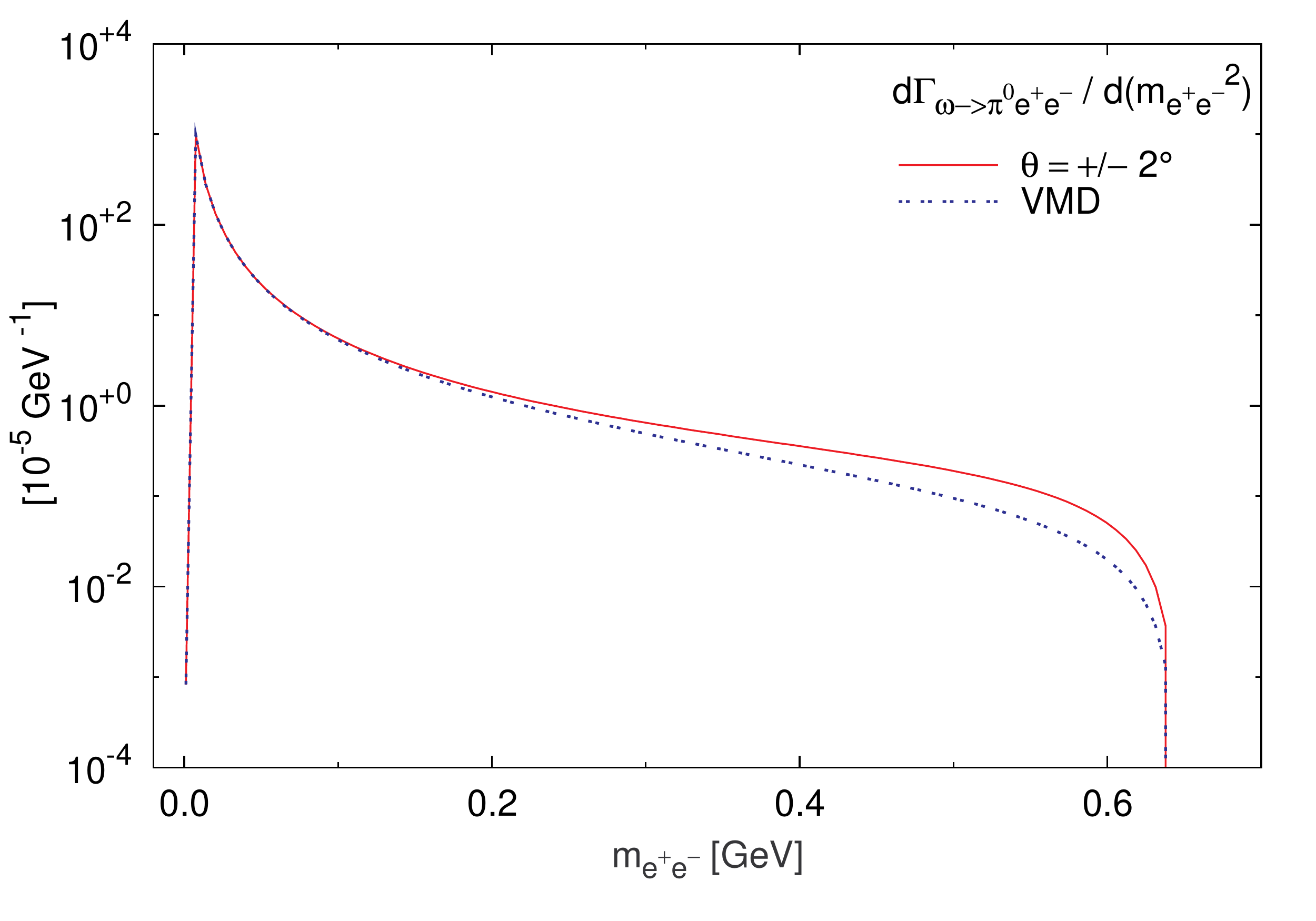} \\
\includegraphics[clip=true,height=0.46\textwidth, angle = -90  ]{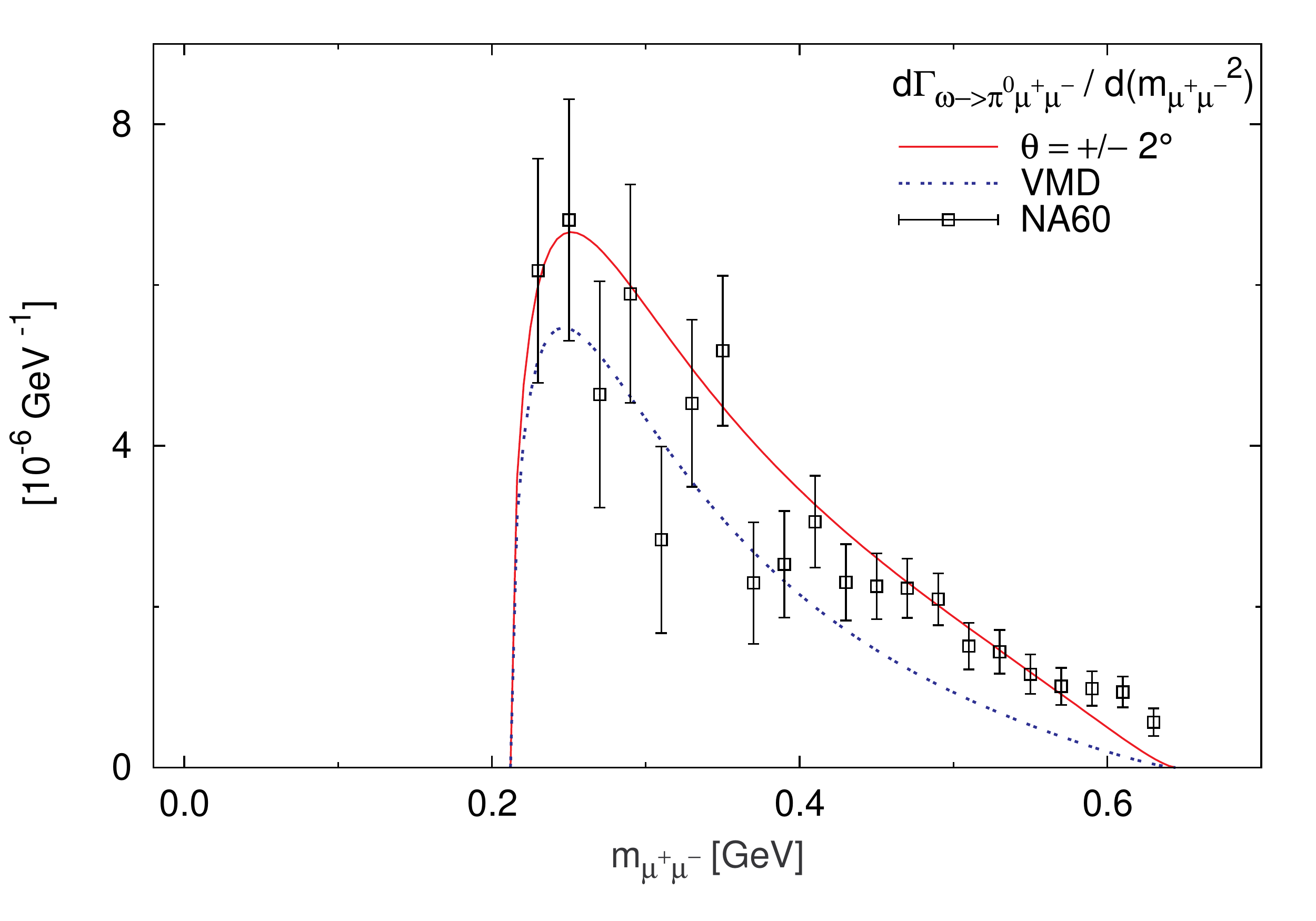}
\caption{Normalized transition form factor of the decay $\omega \rightarrow \pi^0 \,l^+ l^-$  (top).
Single differential decay widths for the decays
$\omega \rightarrow \pi^0 e^+ e^-$ (middle) and $\omega \rightarrow \pi^0 \mu^+ \mu^-$ (bottom).
The solid lines follow with the parameter set (\ref{res-hA}), the dotted lines from the VMD model. The data
have been taken by the NA60 experiment \cite{Arnaldi:2009wb}. }
 \label{fig:ff omega pi}
\end{figure}

In the top panel of Fig.\ \ref{fig:ff omega pi} the normalized form factor of the
$\omega \rightarrow \pi^0\,l^+ l^-$
transition is plotted for the central values of the parameters $h_A$ and $b_A$  as determined from
the photon decays in (\ref{res-hA}). The figure also shows the single-differential decay width for the decays
$\omega \rightarrow \pi^0 \mu^+ \mu^-$ (bottom) and $\omega \rightarrow \pi^0 e^+ e^-$ (middle). 
The solid lines are our leading-order result, the dotted
lines recall the implications of standard vector-meson dominance (VMD)
\begin{eqnarray}
F^{\rm VMD}_{\omega \,\pi^0} = \frac{m_\rho^2}{m_\rho^2-q^2} \,.
\label{def-VMD}
\end{eqnarray}
VMD predictions and possible deviations have also been discussed in \cite{Bando:1993cu,Klingl:1996by}.
The  NA60 data \cite{Arnaldi:2009wb} are described with a $\chi^2/N\simeq 1.8$ in our approach.
This should be compared to the $\chi^2/N\simeq 4.8$ of the VMD model.

We also find a good agreement of the calculated partial decay widths for the decays \cite{Nakamura:2010zzi}
into  dielectrons
\begin{eqnarray}
&& \Gamma_{\omega \rightarrow \pi^0 e^+ e^-}  =  (6.85 \pm 0.21) \cdot 10^{-6} \, {\rm GeV}\,, \qquad
\nonumber\\
&&  \Gamma_{\omega \rightarrow \pi^0 e^+ e^-}^{\rm \,exp} = (6.54 \pm 0.57) \cdot 10^{-6} \, {\rm GeV}\,,
\end{eqnarray}
and dimuons
\begin{eqnarray}
&& \Gamma_{\omega \rightarrow \pi^0 \mu^+ \mu^-}  = (9.74 \pm  0.30) \cdot 10^{-7} \, {\rm GeV}\,,
\nonumber\\
&& \Gamma_{\omega \rightarrow \pi^0 \mu^+ \mu^-}^{\rm \,exp} = (11.04 \pm 3.50) \cdot 10^{-7} \, {\rm GeV}\,,
\end{eqnarray}
where the theoretical error is implied by the parameter variation suggested in (\ref{res-hA}).

As is clearly seen in Fig.\ \ref{fig:ff omega pi} the form factor is probed more sensitively in the
dimuon final state \cite{Landsberg:1986fd}.
In case of the dielectron final state the integral of the differential decay width over the phase space is dominated by kinematical
regions where the form factor has little effect. Therefore the difference of the VMD assumption and our results are more pronounced
in the partial decay width of the $\omega $ meson into the $\pi^0 \mu^+ \mu^-$ final state.

There is no visible difference between the form factor calculated here
and in our previous work \cite{Terschluesen:2010ik}. The slightly different parameter set ($h_A=2.32$, $b_A=0.27$)
of \cite{Terschluesen:2010ik} implies
a $\chi^2/N\simeq 1.7$. The ratio between the term proportional to $b_A$ and the one proportional to $h_A$
in (\ref{eq:f-all}) is
between $0.01$ and $0.02$ in the allowed kinematic region and therefore a change of $b_A$ has little effect on
the $\chi^2$. The differences also would be invisible in Fig.\ \ref{fig:ff omega pi}.

We find it an encouraging result that our power counting as formulated first in  \cite{Lutz:2008km} predicted the relevance of
precisely those terms in the effective Lagrangian that later \cite{Terschluesen:2010ik} lead
to a quantitative description of the NA60 data \cite{Arnaldi:2009wb}.

\subsection{Decay $\omega \rightarrow \eta \,l^+ l^-$}

\begin{figure}[h!]
\center
\includegraphics[clip=true,width=0.3\textwidth]{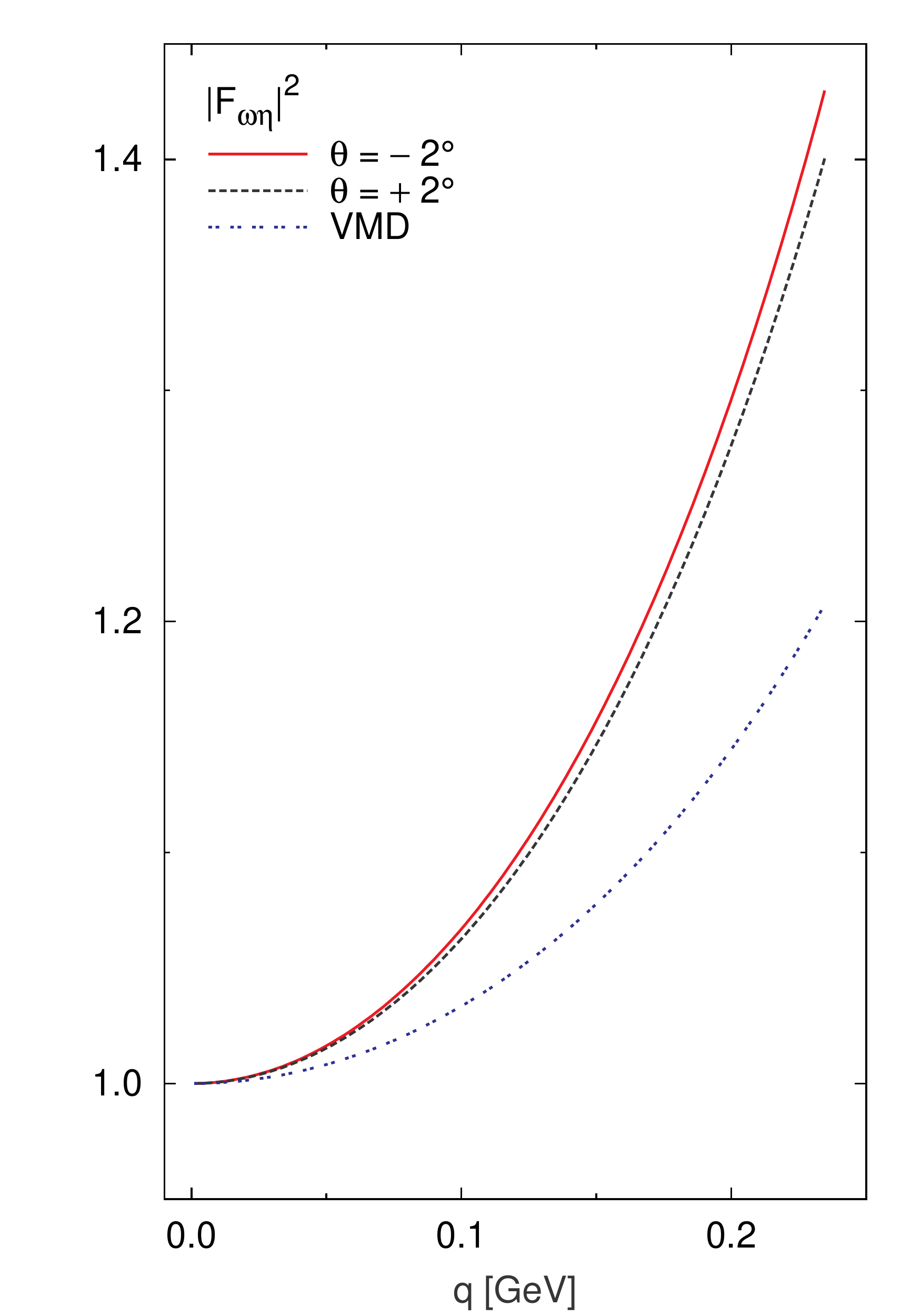} \\
\includegraphics[clip=true,height=0.46\textwidth, angle = -90]{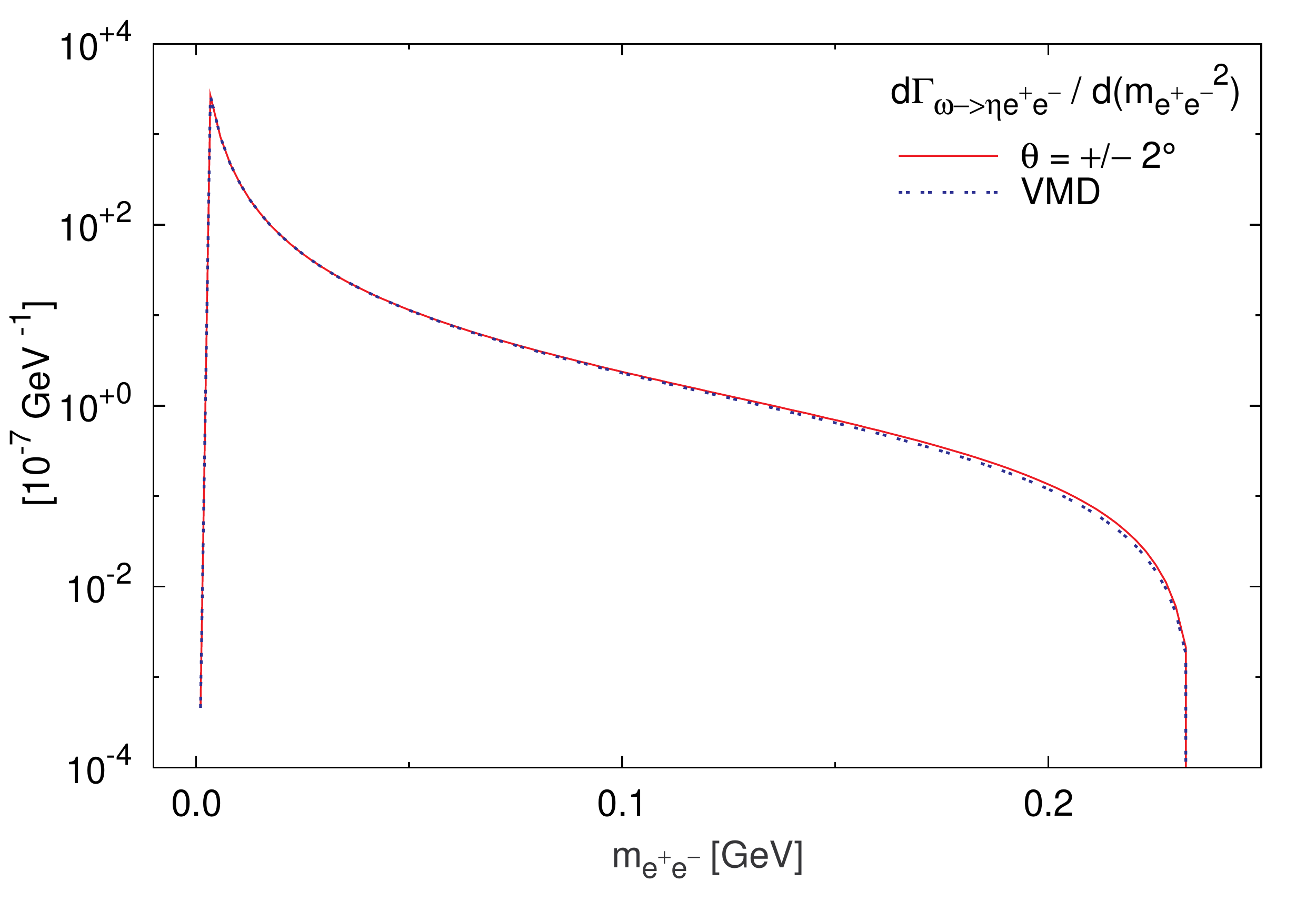} \\
\includegraphics[clip=true,height=0.46\textwidth, angle = -90  ]{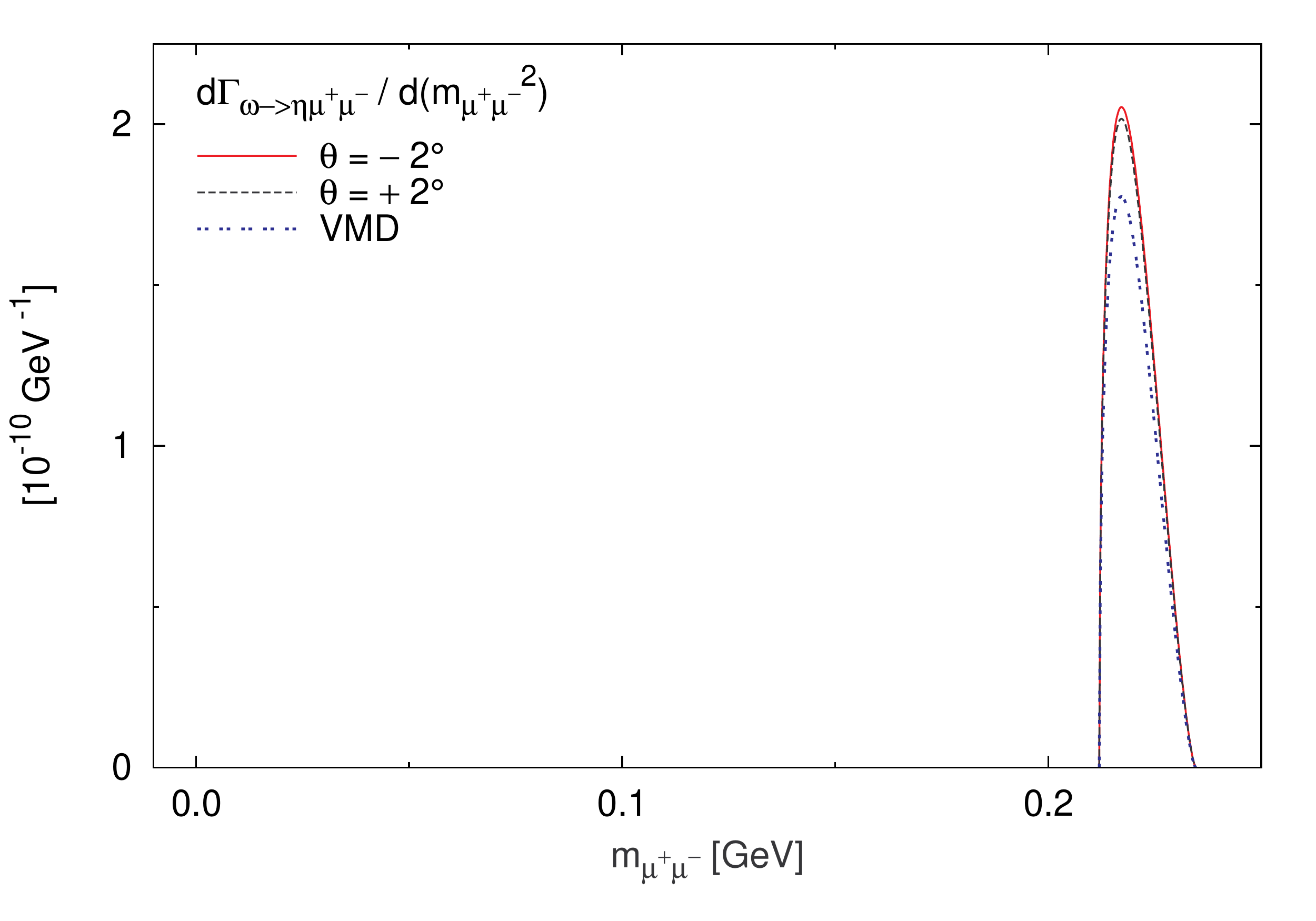}
\caption{Normalized transition form factor of the decay $\omega \rightarrow \eta \,l^+ l^-$  (top).
Single differential decay widths for the decays
$\omega \rightarrow \eta e^+ e^-$  (middle) and $\omega \rightarrow \eta \mu^+ \mu^-$ (bottom).
The solid and dashed lines show our predictions using the central values of (\ref{res-hA}, \ref{res-eH})
and the variation implied by (\ref{f0-range}). The dotted lines display the VMD result (\ref{eq:omega-eta-VMD}). }
 \label{fig:omega-eta}
\end{figure}

In Fig.\ \ref{fig:omega-eta} the normalized form factor for $\omega \rightarrow \eta\, l^+ l^-$ and the differential
decays into dimuons and dielectrons are plotted. The solid and dashed lines show our predictions for negative or
positive mixing angle, respectively. They are compared to the dotted lines which
are implied by the VMD picture.  As the transition $\omega \rightarrow \eta \,l^+ l^-$ can only happen via a virtual
$\omega$ meson (assuming OZI suppression), the VMD form factor equals
\begin{align}
F_{\omega \,\eta}^{\rm VMD} (q^2) = \frac{m_\omega^2}{m_\omega^2 - q^2}\, .
\label{eq:omega-eta-VMD}
\end{align}
In contrast to the transition $\omega \rightarrow \pi^0 \, l^+ l^-$ the form factor
for the $\omega \rightarrow \eta \, l^+ l^-$  transition depends on the parameters $h_H$, $e_{H}$ and $f_H$. However, given the smallness of
our mixing angle in (\ref{res-hA}) the influence of those parameters is quite small.
According to (\ref{eq:f-all}, \ref{eq:f-mixing})
only the singlet amplitude $f_{\omega \,\eta_1}$ depends on $h_H$, $e_{H}$ and $f_H$. The small widening of the solid lines are implied
by a variation of the parameter $f_H$ with $f \leq f_H \leq \sqrt{2}\,f$. Fig.\ \ref{fig:omega-eta} illustrates a significant deviation of
our predictions from the standard VMD picture. While the form factor and the differential
decays into dimuons is highly discriminative, the dielectron final state appears less sensitive to the details of the transition form factor.

We predict the following partial decay widths
\begin{eqnarray}
&& \Gamma_{\omega \rightarrow \eta \,e^+ e^-} = (2.88 \pm 0.22) \cdot 10^{-8} \, {\rm GeV} \,,
\nonumber\\
&&\Gamma_{\omega \rightarrow \eta \,\mu^+ \mu^-} = ( 8.57 \pm 0.64) \cdot 10^{-12} \, {\rm GeV} \,,
\end{eqnarray}
where we estimated the theoretical error by a variation of our parameters according to (\ref{res-hA}, \ref{res-eH}, \ref{f0-range}).
As there are no experimental data available for the decays into dileptons, our calculations must be seen
as predictions. For the corresponding branching ratios we predict
\begin{eqnarray}
&& {\rm Br}_{\omega \rightarrow \eta \,e^+ e^-} = (3.39 \pm 0.26) \cdot 10^{-6}  \,,
\nonumber\\
&& {\rm Br}_{\omega \rightarrow \eta \,\mu^+ \mu^-} = ( 1.01 \pm 0.08) \cdot 10^{-9} \,.
\end{eqnarray}

\subsection{Decay $\phi \rightarrow \eta \,l^+ l^-$}

\begin{figure}[h!]
\center
\includegraphics[clip=true,width=0.3\textwidth]{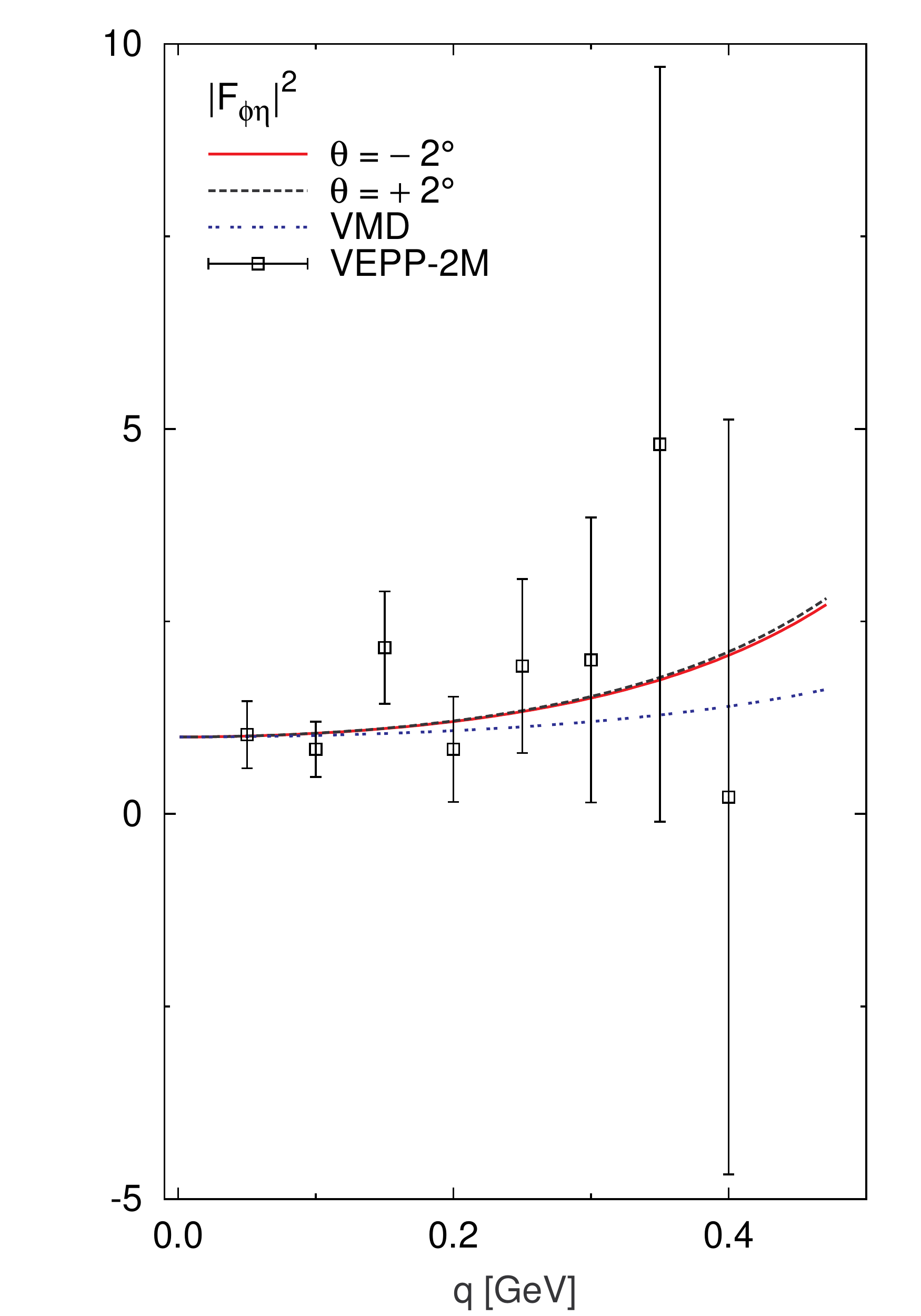} \\
\includegraphics[clip=true,height=0.46\textwidth, angle = -90]{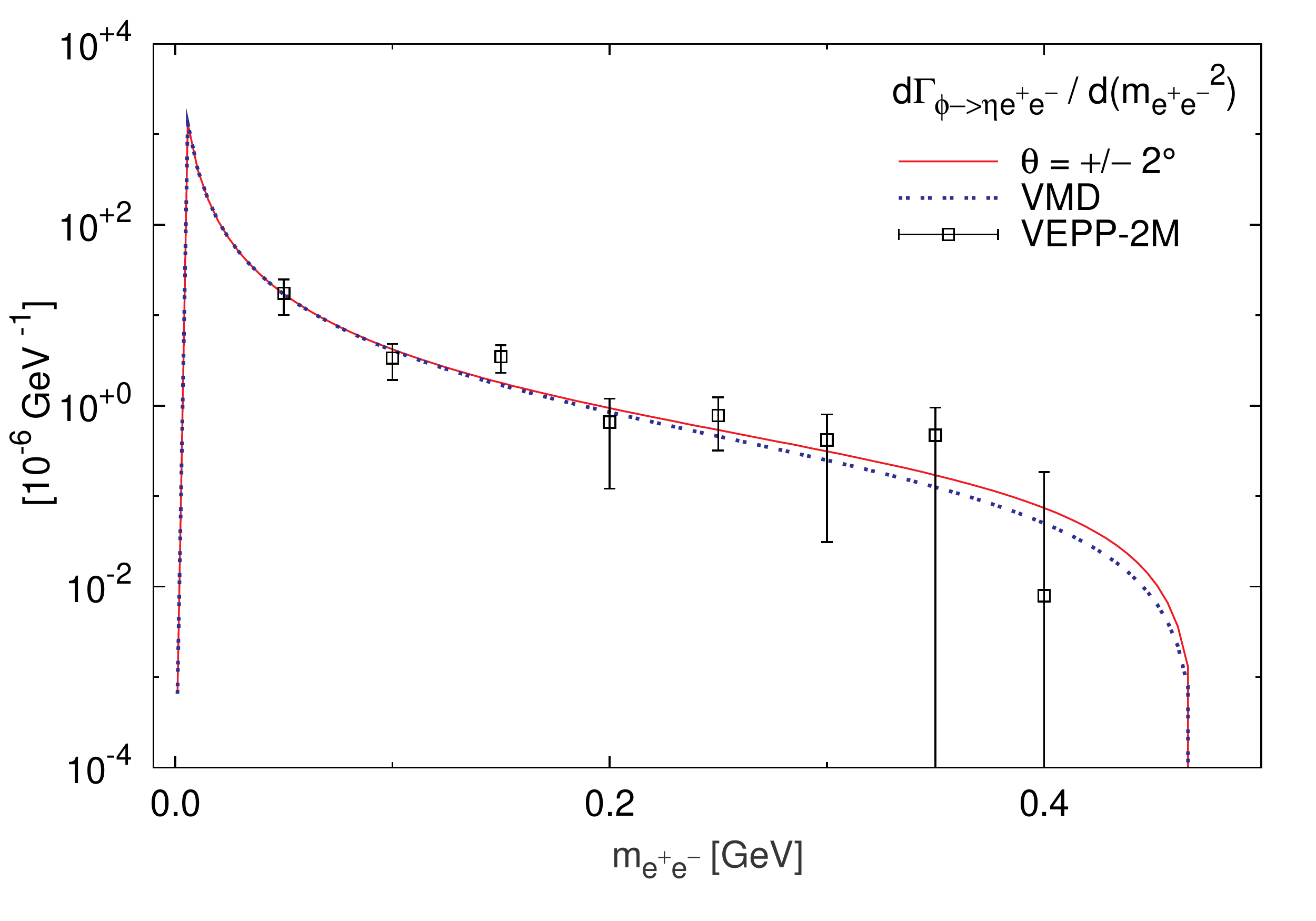} \\
\includegraphics[clip=true,height=0.46\textwidth, angle = -90  ]{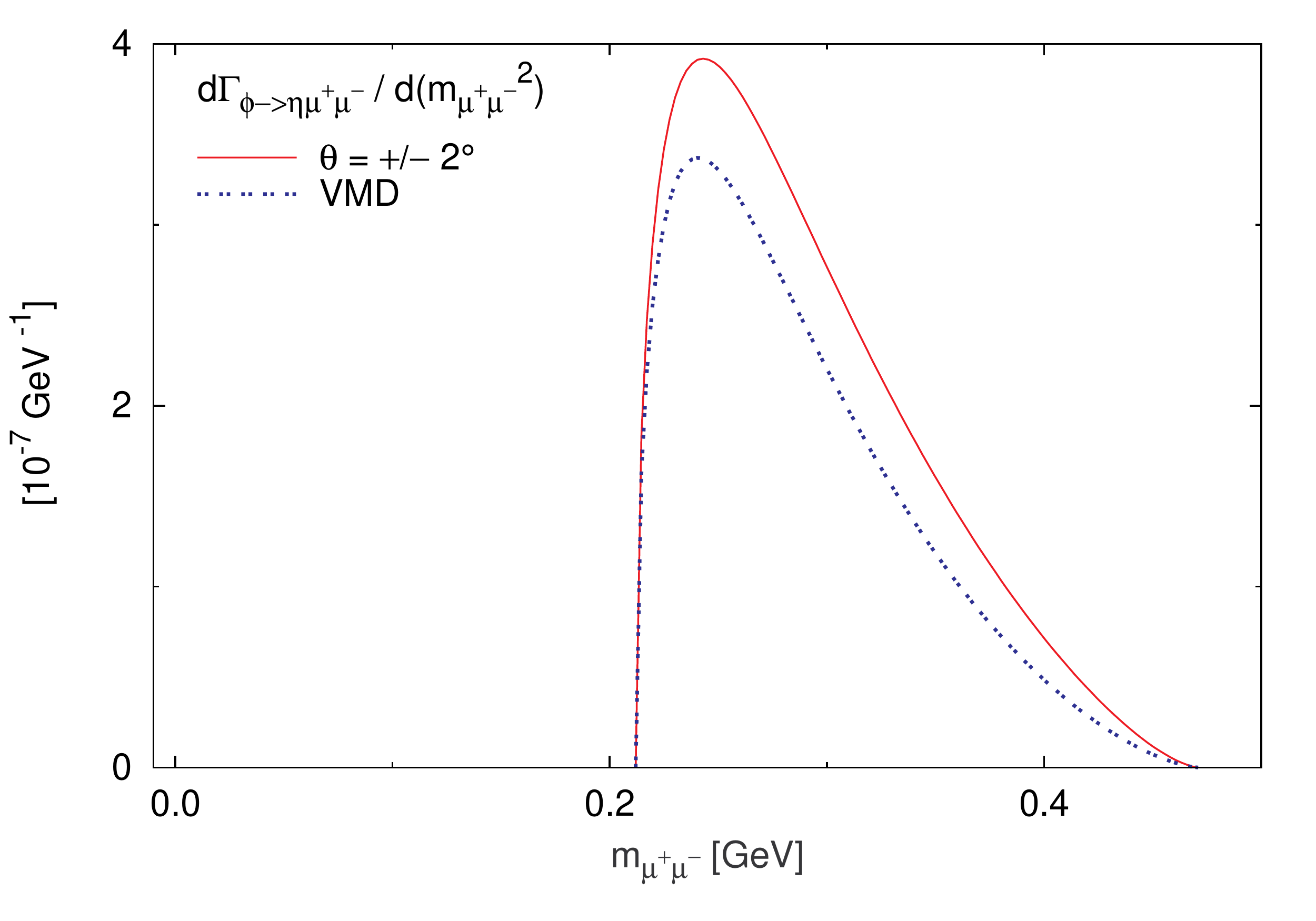}
\caption{Normalized transition form factor of the decay $\phi \rightarrow \eta \,l^+ l^-$  (top).
Single differential decay widths for the decays
$\phi \rightarrow \eta e^+ e^-$ (middle) and $\phi \rightarrow \eta \mu^+ \mu^-$ (bottom).
The solid and dashed lines show our predictions using the central values of (\ref{res-hA}, \ref{res-eH})
and the variation implied by (\ref{f0-range}). The dotted lines display the VMD result (\ref{eq:phi-eta-VMD}).
The data have been taken at the VEPP-2M collider \cite{Achasov:2000ne}.}
 \label{fig:phi-eta}
\end{figure}

In Fig.\ \ref{fig:phi-eta} the normalized form factor for $\phi \rightarrow \eta\, l^+ l^-$ and the differential
decays into dimuons and dielectrons are plotted. The solid and dashed lines show our predictions for negative or
positive mixing angle, respectively. They are compared to the dotted lines which
are implied by the VMD picture.  As the transition $\phi \rightarrow \eta \,l^+ l^-$ can only happen via a virtual
$\phi$ meson (assuming OZI suppression), the VMD form factor equals
\begin{align}
F_{\phi \,\eta}^{\rm VMD} (q^2) = \frac{m_\phi^2}{m_\phi^2 - q^2}\, .
\label{eq:phi-eta-VMD}
\end{align}
A deviation from standard VMD is visible in the form factor and the single-differential decay widths for the
decays into dimuon and dielectron. Like in the $\omega \to \eta \,l^+l^-$ decay the uncertainty of our
prediction is quite small, at least the one emerging from the variation of $f_H$ as given in (\ref{f0-range}).
This is because the form factor is dominated by its octet component, which does not depend on the
parameters $h_H, e_H$ and $f_H$.

The calculated form factor is compared to data taken at the VEPP-2M collider \cite{Achasov:2000ne} for
the decay $\phi \rightarrow \eta\, e^+ e^-$. As the error bars of the data are relatively large they are not
very discriminative. In the near future data with much smaller error bars are expected from the KLOE collaboration.

Integrating the single-differential widths for the decays into a dielectron or a dimuon yields the values
\begin{eqnarray}
&& \Gamma_{\phi \to \eta\, e^+ e^-\,} = (4.81 \pm 0.59) \cdot 10^{-7} \, {\rm GeV}\,,
\nonumber\\
&& \Gamma_{\phi \to \eta \,\mu^+ \mu^-} = ( 2.83 \pm 0.33) \cdot 10^{-8} \, {\rm GeV}\,,
\end{eqnarray}
with uncertainties estimated according to (\ref{res-hA}, \ref{res-eH}, \ref{f0-range}).
Our results are in good agreement with the experimental constraints \cite{Nakamura:2010zzi}
\begin{eqnarray}
&& \Gamma_{\phi \to \eta \,e^+ e^-\,}^{\rm exp.} = \left( 4.90 \pm 0.47 \right) \cdot 10^{-7} \, {\rm GeV}\,,
\nonumber\\
&& \Gamma_{\phi \to \eta \,\mu^+ \mu^-}^{\rm exp.} < 4.00 \cdot 10^{-8} \, {\rm GeV}\,.
\label{eq:expphieta}
\end{eqnarray}

Given our small mixing angle $\theta$, according to (\ref{res-hA}), the results of the three so far considered
electromagnetic transitions $\omega \to \pi^0$, $\omega \to \eta$ and $\phi \to \eta$ do not deviate much
from our previous analysis \cite{Terschluesen:2010ik}, where the $\eta$ meson has been treated as a pure octet state.
In the following, we will study decays which involve the $\etapr$. Such processes have not been studied so far
with our chiral Lagrangian since the inclusion of the pseudoscalar singlet has only been achieved in the present
work.

\subsection{Decay $\phi \rightarrow \etapr e^+ e^-$}

\begin{figure}[h]
\center
\includegraphics[clip=true,width=0.35\textwidth]{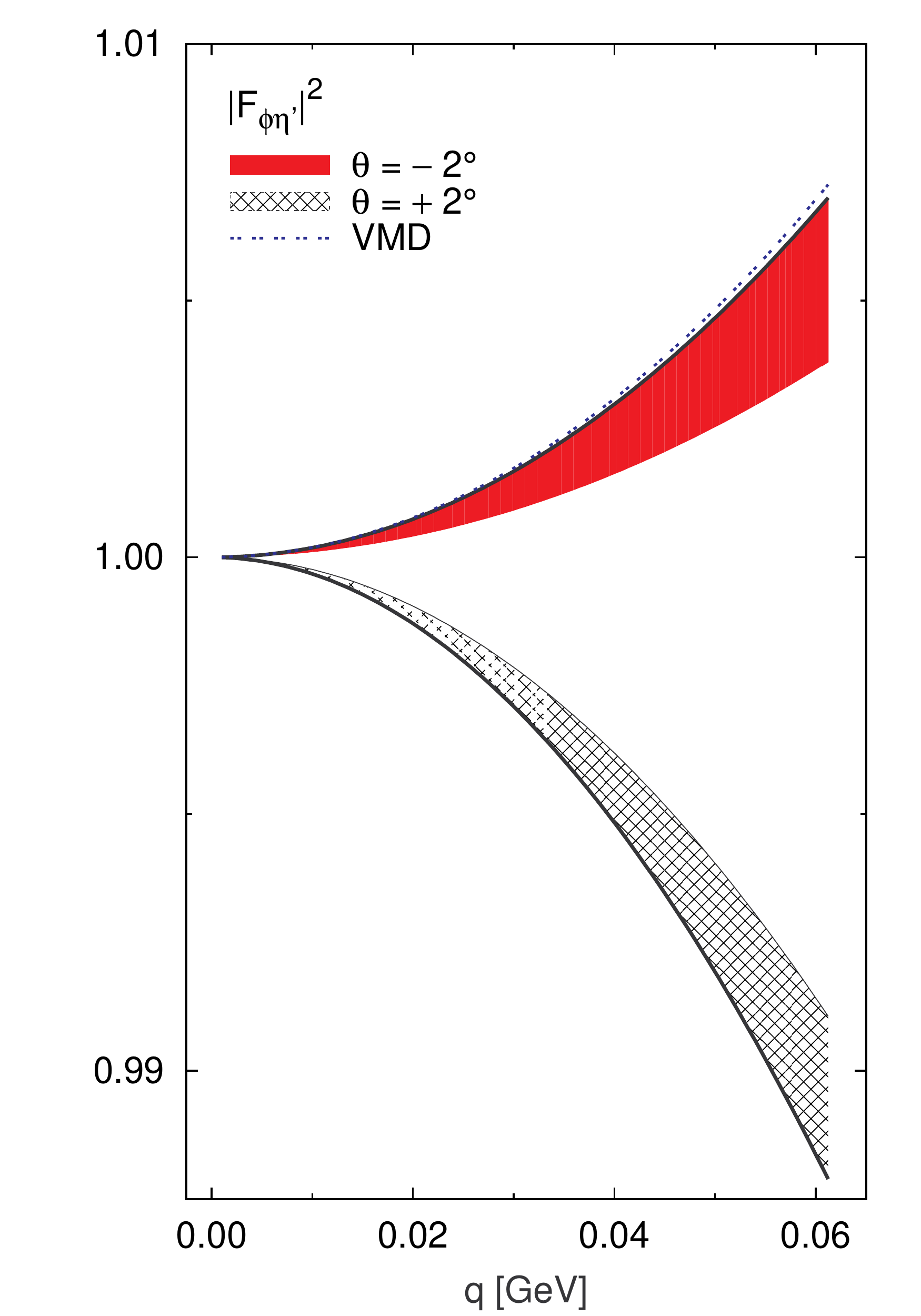}
\includegraphics[clip=true,height=0.47\textwidth, angle = -90]{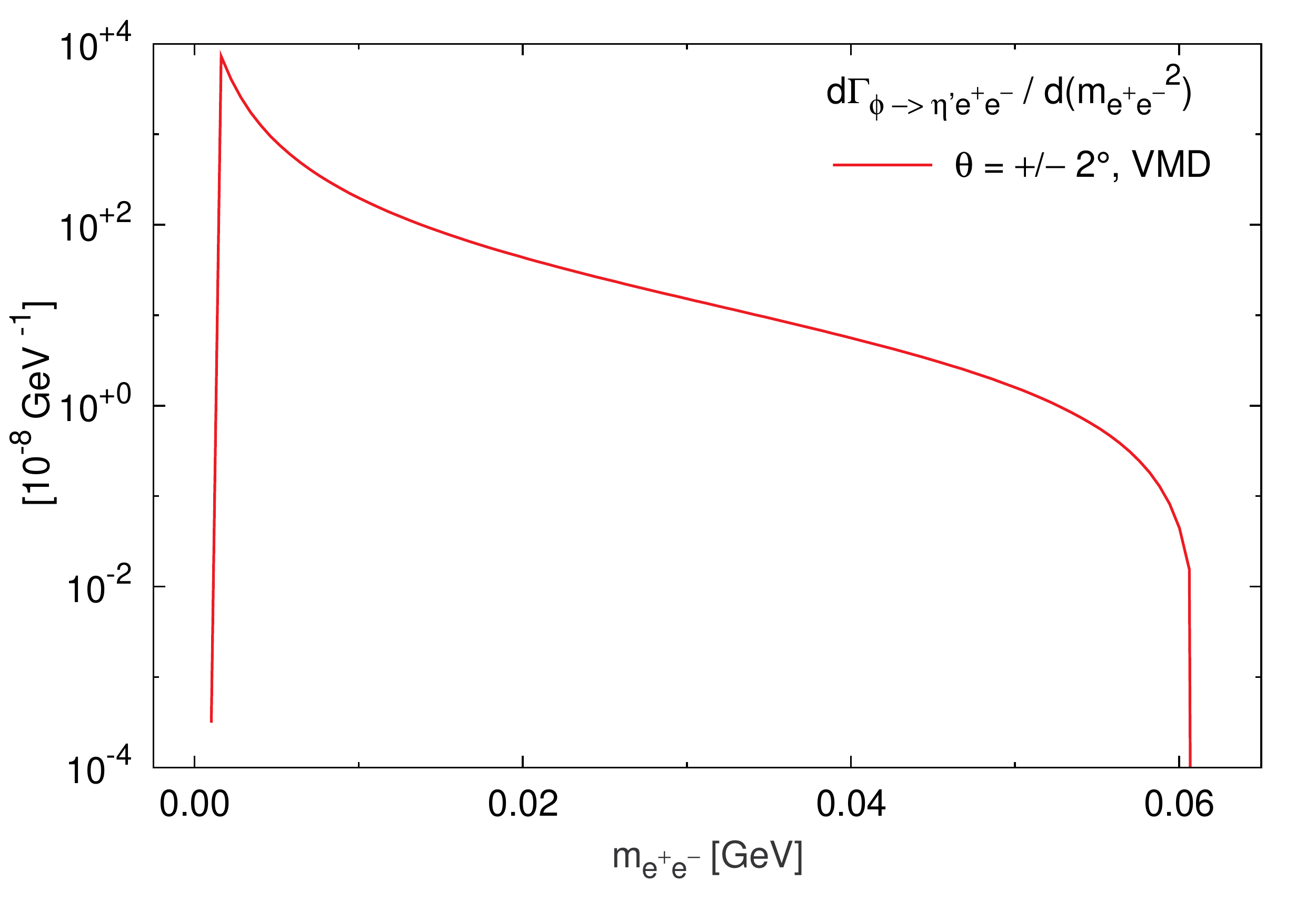}
\caption{Normalized form factor for the $\phi \rightarrow \etapr$ transition (top)
and single differential decay width $\phi \rightarrow \etapr e^+ e^-$ (bottom). The bands are implied by the
central values of the two parameter sets
(\ref{res-hA}, \ref{res-eH}) and the range (\ref{f0-range}). The dotted line is the VMD prediction
(\ref{eq:phi-eta-VMD}). All lines coincide for the single differential decay rate.}
\label{fig:phi-etaprime}
\end{figure}

The form factor for the $\phi \rightarrow \etapr \,e^+e^-$ transition tests our counting scheme in the $\etapr$ sector.
The two parameters $h_H$ and $e_H$ introduced in the hadrogenesis Lagrangian (\ref{def-L3})
have a significant impact on the associated photon decay $\phi \rightarrow \etapr \,\gamma$.
In Fig. \ref{fig:phi-etaprime}
we present our prediction for the form factor and the single-differential decay width for the decay into a dielectron.
As the upper limit of the allowed kinematic region, $m_\phi - m_{\etapr} = 61 \, {\rm MeV}$, is smaller than the mass of a dimuon,
$2 \,m_\mu = 212 \, {\rm MeV}$, a decay into a dimuon is not possible.

Using OZI suppression, which is incorporated in our leading-order Lagrangian, a decay of a $\phi$ meson
into one of the isospin-singlet $\eta$ states can only happen via a virtual $\phi$-meson.
Correspondingly, the standard VMD form factor depends on the $\phi$-meson mass only and is the same as for the $\phi$-meson decay into
an $\eta$ meson (\ref{eq:phi-eta-VMD}). Owing to the existence of two distinct solutions (\ref{res-eH}) the form factor
in Fig.\ \ref{fig:phi-etaprime} shows two branches. Both branches differ from the implications of the VMD picture.
Our scenario with a negative mixing angle shows a significantly smaller difference. The thick boundary lines
of the two branches give
the results with the particular choice $f_H =f$. Since the form factor does not significantly deviate from 1
in the allowed kinematic
range all the distinct curves in the top panel of Fig.\ \ref{fig:phi-etaprime}
lead to the same line for the single-differential decay rate. In view of the very small deviations of
the form factor from unity an experimental discrimination of the
presented scenarios is extremely challenging for this decay process.
The integrated decay width is given by
\begin{align}
 \Gamma_{\phi \to \etapr e^+ e^-}  = (1.39 \pm 0.30) \cdot 10^{-9} \, {\rm GeV}  \,.
\end{align}
The corresponding branching ratio is
\begin{align}
 {\rm Br}_{\phi \to \etapr e^+ e^-}  = (3.25 \pm 0.70) \cdot 10^{-7} \,.
\end{align}
Since there are no experimental data available, our
result has to be seen as a prediction. As there is no significant dependence on the form factor, however, one
merely would probe the prediction from quantum electrodynamics. The situation is significantly better for the
transition of the $\etapr$ to the $\omega$, to which we turn next.

\subsection{Decay $\etapr \rightarrow \omega \,e^+ e^-$}

\begin{figure}[h]
\center
\includegraphics[clip=true,width=0.35\textwidth]{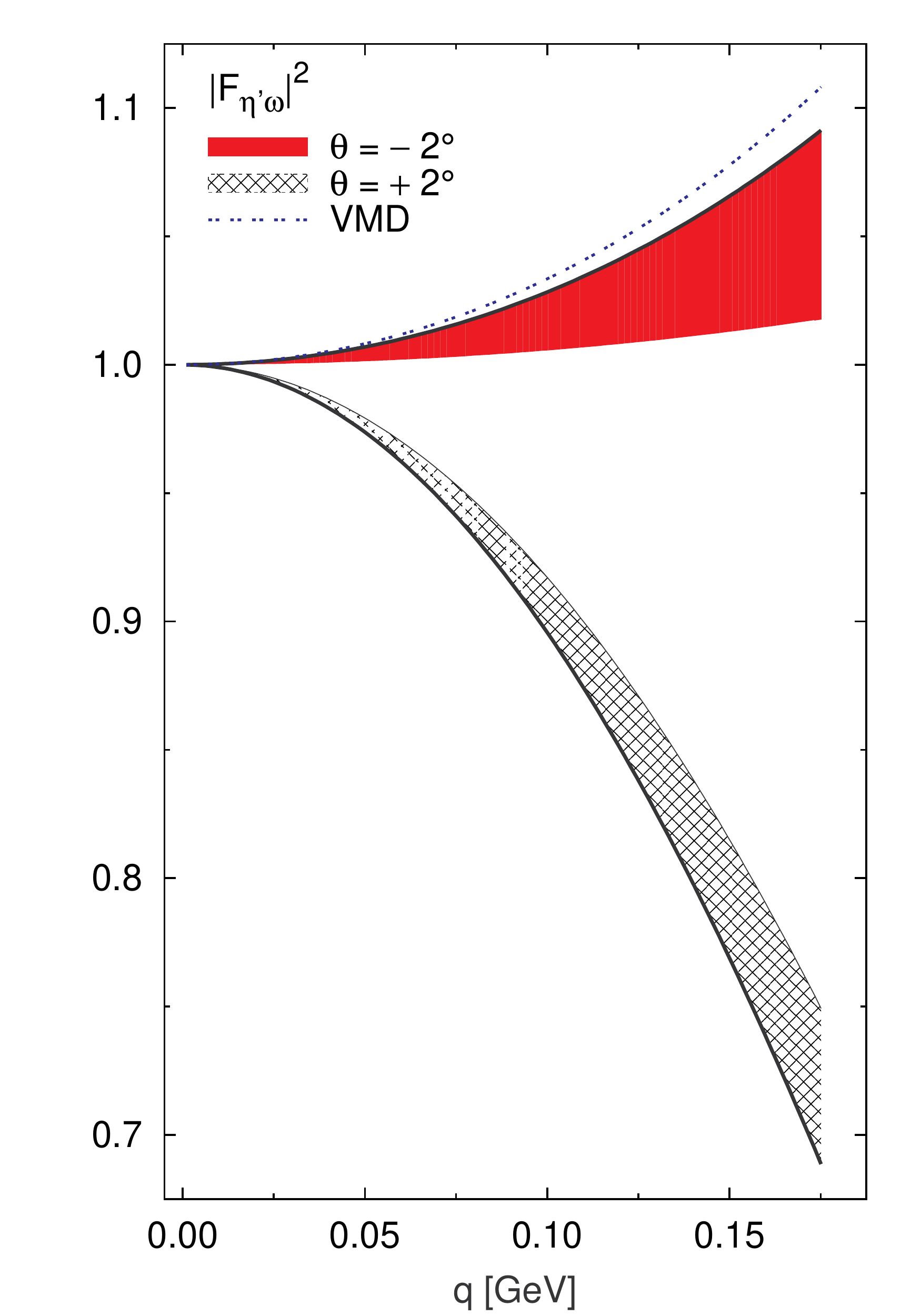}
\includegraphics[clip=true,height=0.47\textwidth, angle = -90]{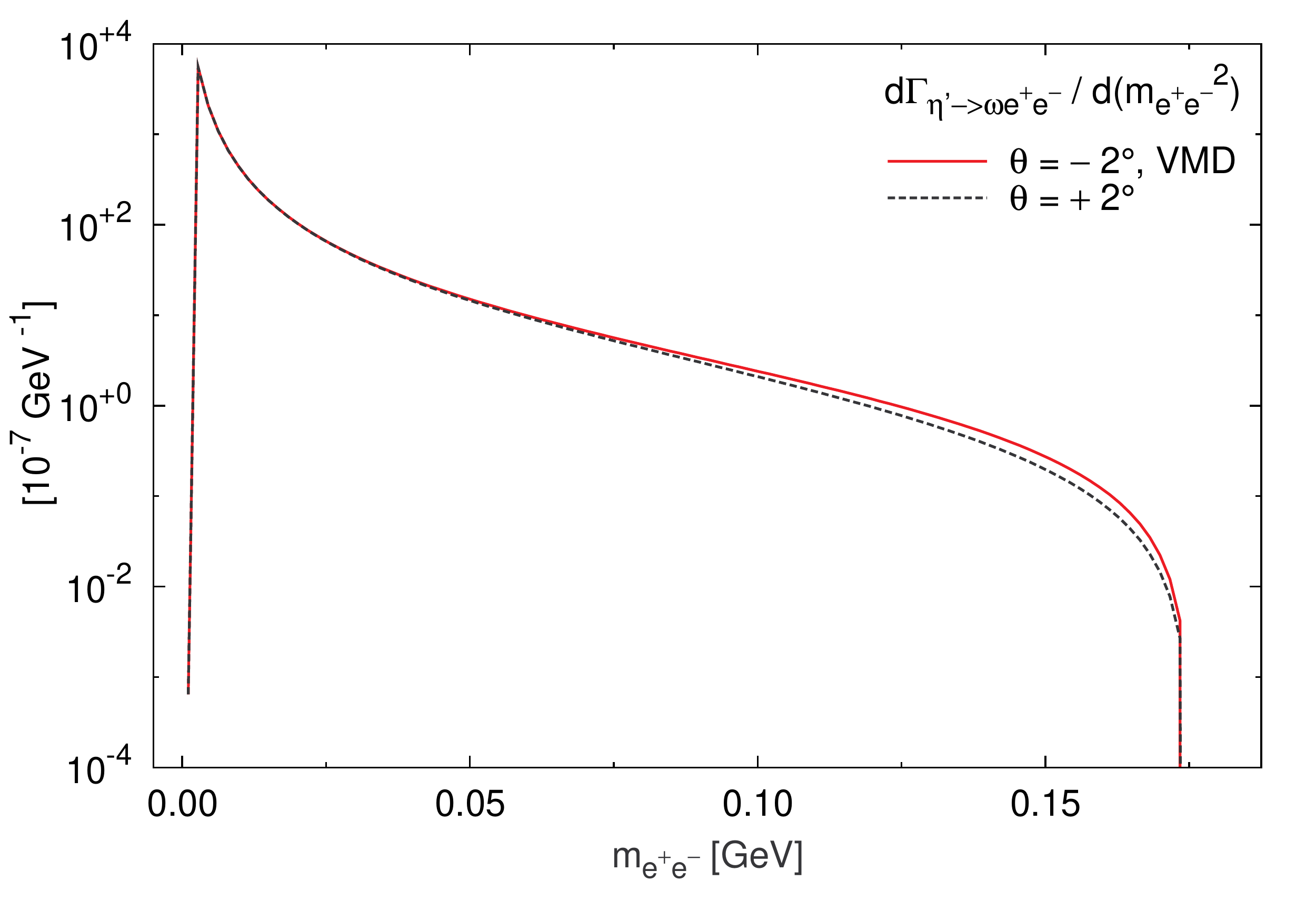}
\caption{Normalized form factor for the $\etapr \rightarrow \omega$
transition (top) and single differential decay rate for $\etapr \rightarrow \omega\, e^+ e^-$ (bottom).
The bands are implied by the central values of the two parameter sets
(\ref{res-hA}, \ref{res-eH}) and the range (\ref{f0-range}).
The dotted line is the VMD prediction (\ref{eq:etaprime-omega-VMD}).}
\label{fig:etaprime-omega}
\end{figure}

A stringent test of our power counting is provided by the decay of the $\etapr$ into $\omega \,e^+ e^-$.
Due to kinematic reasons there is no decay into a dimuon possible. The normalized form factor  and the
single-differential decay rate in comparison to the VMD picture are plotted in Fig.\ \ref{fig:etaprime-omega}.
On account of the OZI suppression the vector VMD form factor is
\begin{eqnarray}
F_{\etapr \,\omega}^{\rm VMD}(q^2) = \frac{m_\omega^2}{m_\omega^2-q^2} \,.
\label{eq:etaprime-omega-VMD}
\end{eqnarray}
For this decay the uncertainties in the parameters $f_H$ yield a significant uncertainty
of our prediction visualized by the bands in Fig.\ \ref{fig:etaprime-omega}.
For all parameter choices, however, the form factor shows a significant difference to the
VMD implications. The two branches in the
form factor reflect the two distinct parameters sets (\ref{res-eH}). The negative mixing-angle solution
predicts a form factor that
is closer to the VMD result, in particular for the choice of a small $f_H \approx f$. This choice is highlighted by
the thick lines at the edges of the bands. A precise measurement of the from factor would help to determine
the wave-function normalization of the $\etapr$ meson, i.e.\ the magnitude of the $f_H$ parameter.
Possibly the BES-II experiment might have enough statistics on $\etapr$ decays to determine the transition
form factor of $\etapr \to \omega$.

For the integrated width we predict
\begin{eqnarray}
&& \Gamma_{\etapr \to \omega \,e^+ e^-} =  ( 3.36 \pm 1.10 ) \cdot 10^{-8} \, {\rm GeV}\,,
\end{eqnarray}
with a theoretical error as implied by (\ref{res-hA}, \ref{res-eH}, \ref{f0-range}). This leads to a branching
fraction of
\begin{eqnarray}
&& {\rm Br}_{\etapr \to \omega \,e^+ e^-} =  ( 1.69 \pm 0.56 ) \cdot 10^{-4}  \,.
\end{eqnarray}
Note that the corresponding prediction of
the VMD picture is within our error bars. While the integrated width is not very sensitive to the
unknown size of $f_H$, a precise measurement of the differential width at dielectron masses around 140 MeV
could be discriminating.

\section{Summary} \label{sec:summary}

We have constructed a chiral Lagrangian with vector-meson fields and the $\etapr$ field. A generalized counting
scheme has been introduced that considers the nonets of vector and pseudoscalar degrees of freedom as light.
It has been argued that such an assumption can be justified with the hadrogenesis conjecture,
which would imply a significant mass gap in the meson spectrum of QCD in
the large-$N_c$ limit. In application of our counting scheme we constructed the complete
leading-order Lagrangian, where an explicit realization of the underlying naturalness assumption was provided.
Our Lagrangian systematizes and  extends our previous studies that did not involve the $\etapr$ field \cite{Lutz:2008km,Leupold:2008bp,Terschluesen:2010ik}. 

We do not claim that we have already a systematic power counting or that the suggested counting rules can by justified by 
large-$N_c$ arguments.
Clearly, a detailed study of loop effects is necessary before such a claim can be made. The logic of our work is not that 
large-$N_c$ can justify our counting rules, rather we formulate our dynamical assumption in the large-$N_c$ context of QCD. 
The conjectured scale separation in the large-$N_c$ meson spectrum is motivated by 
phenomenology \cite{Lutz:2001dr,Lutz:2003fm,Lutz:2004dz,Lutz:2005ip,Lutz:2007sk}.

As an application of our leading-order Lagrangian we studied the electromagnetic transitions
of vector to pseudoscalar mesons. Adjusting the five relevant leading-order parameters we have reproduced the
empirical decay rates of $\omega \to \pi^0\,\gamma, \, \eta \,\gamma$, of
$\phi \to \eta\, \, \gamma, \etapr\, \gamma$ and of
$\etapr \to \omega\,\gamma$ for real photons in the respective final state.
As a striking consequence we have found an unconventionally small mixing angle
$\theta \simeq \pm 2^\circ$ of the $\eta$-$\etapr$ system. Phenomenological analyses prefer a two-mixing-angle
scenario with significantly larger mixing angles \cite{Feldmann:1998vh,Feldmann:1998sh,Escribano:2005qq}. Going
beyond leading order in our scheme provides us with the possibility of two mixing angles.
It remains to be seen whether such a full next-to-leading-order treatment including loops leads to a sizable change
of the mixing angle(s) and/or adds new aspects to the phenomenological analyses.

Given our parameter set we predicted the electromagnetic transitions
of vector to pseudoscalar mesons.
For all electromagnetic transition form
factors we find significant deviations from the expected behavior of the vector-meson dominance (VMD) model.
As in \cite{Terschluesen:2010ik} the description of the data for the $\omega \rightarrow \pi^0$ transition
form factor measured by the NA60
collaboration \cite{Arnaldi:2009wb} is much better than the description with the standard VMD form factor.
Furthermore, for all decays the branching ratios of the decay widths into dileptons agree very well with the available
experimental ratios. Quantitative predictions for the so far unknown decay rates
$\omega \to \eta \,e^+e^-$, $\omega \to \eta \,\mu^+\mu^-$,
$\phi\to \eta\,\mu^+ \mu^-  $, $\phi\to \etapr\,e^+ e^-  $ and $\etapr \to \omega \,e^+ e^-$ and the corresponding
form factors have been provided.

\begin{appendix}

\section{Parameter determination for a fixed $\eta$-$\etapr$ mixing angle}
\label{sec:appendix}

As pointed out in subsection \ref{subsec:naturalness assumption}, the leading-order contributions to the
two-point functions of the pseudoscalar mesons \eqref{eq:L2pseudo} can be used to determine the
$\eta$-$\etapr$ mixing angle $\theta$ (in leading order) as a function of the empirical $\eta$ and $\etapr$
masses yielding $\theta \approx -11^\circ$. Naively one might think that, if the mixing angle is fixed, the
empirical matrix elements of the five two-body decays $\omega \rightarrow \pi^0 \gamma$,
$\omega \rightarrow \eta \gamma$, $\etapr \rightarrow \omega \gamma$, $\phi \rightarrow \eta \gamma$ and
$\phi \rightarrow \etapr \gamma$ can be used to determine the five open parameters $h_A$, $b_A$, $e_H$, $h_H$ and
$f_H$ needed to calculate the radiative decays in section \ref{sec:Lagrangian}.
However, this is not the case. For a given mixing angle, decays including the $\eta$ or $\etapr$
meson can be rewritten into formal
decays for the octet and singlet states $\eta_8$ and $\eta_1$, respectively. These formal decay amplitudes
are defined as (cf.\ \eqref{eq:f-mixing})
\begin{eqnarray}
  && f_{\omega \, \eta_8} = f_{\omega \,\eta} \, \cos\theta + f_{\omega  \,\etapr} \, \sin\theta \,, 
  \nonumber \\[0.3em] 
  && f_{\phi \, \eta_8} = f_{\phi \,\eta} \, \cos\theta + f_{\phi  \,\etapr} \, \sin\theta \,,
  \nonumber \\
  && f_{\omega \, \eta_1} = \frac{f_H}{f} \, \left(
    f_{\omega \,\etapr} \, \cos\theta - f_{\omega  \,\eta} \, \sin\theta
  \right) \,,  
  \nonumber \\
  && f_{\phi \, \eta_1} = \frac{f_H}{f} \, \left(
    f_{\phi \,\etapr} \, \cos\theta - f_{\phi  \,\eta} \, \sin\theta
  \right) \,.
  \label{eq:formalapp}
\end{eqnarray}
One gets three formal decays not involving the singlet state, $\omega \rightarrow \pi^0 \gamma$,
$\omega \rightarrow \eta_8 \gamma$ and $\phi \rightarrow \eta_8 \gamma$, and two involving only the singlet
state, $\omega \rightarrow \eta_1 \gamma$ and $\phi \rightarrow \eta_1 \gamma$.

The strategy is to use the theoretical formulae \eqref{eq:f-all} for the left hand side of \eqref{eq:formalapp}
(evaluated for the photon point $q^2=0$) and the experimental numbers of \eqref{empirical-fs} for the right hand side.
The \emph{three} decays not
involving the singlet state depend only on the \emph{two} parameters $h_A$ and $b_A$ and are used to determine those.
Of course, the signs of the amplitudes on the right hand side of \eqref{eq:formalapp} are not determined
by the respective decay widths, see \eqref{empirical-fs}.
Yet, it is not possible to fit all three decays in good agreement with the experimental data. For all possible
sign combinations for the amplitudes, fitting of $h_A$ and $b_A$ yields a minimal
$\chi^2$ of $\chi^2(\theta = -11^\circ) = 25.6$. Larger absolute values of $\theta$ yield even worse results,
$\chi^2(\theta = -15^\circ) = 40.9$  and $\chi^2(\theta = -20^\circ) = 61.3\,$.\footnote{In \cite{Gasser:1984gg},
next-to-leading-order effects are used to determine an $\eta$-$\etapr$ mixing angle of $\theta = -20^\circ$.}
In view of these facts, we decided to use the mixing angle $\theta$ as an additional parameter to describe the
three decays not involving the singlet state. This is discussed in the main text.

Additionally, one has only two decays involving the singlet state $\eta_1$ which has to be used to determine the
remaining three parameters $e_H$, $h_H$ and $f_H$. Thus, one needs additional data to fix all parameters.
In the main text we have determined all parameters except for $f_H$.
\end{appendix}

\bibliography{literature-t}{}

\begin{thebibliography}{10}
\bibitem{Zweig:1981pd}
G.~Zweig, {An SU(3) model for strong interaction symmetry and its breaking},
  CERN-TH-401.
\end{thebibliography}
\bibliographystyle{elsart-num}

\end{document}